\documentclass[12pt,preprint]{aastex}

\usepackage{graphicx}
\usepackage{multicol}
\usepackage{wrapfig}
\usepackage{natbib}

\newcommand{\be}{\begin{equation}}
\newcommand{\ee}{\end{equation}}
\newcommand{\nn}{\mbox{} \nonumber \\ \mbox{} }
\newcommand{\ba}{\begin{eqnarray}}
\newcommand{\ea}{\end{eqnarray}}
\newcommand{\om}{\omega}

\renewcommand{\v}{{\bf v}}

\newcommand\eg{\textit{e.g.\ }}

\newcommand{\Bf}{{magnetic field}}
\newcommand{\Bfs}{{magnetic fields}}
\newcommand{\Ef}{{electric  field}}

\newcommand{\NS}{neutron star}
\newcommand{\NSs}{{neutron stars}}

\newcommand{\ms}{magnetosphere}
\newcommand{\mss}{magnetospheres}

\newcommand{\Lf}{Lorentz factor}
\newcommand{\dotE}{\dot{E}}

\begin{document}

\title{Fast radio bursts as giant pulses from young rapidly rotating pulsars}
\author{Maxim Lyutikov$^1$, Lukasz Burzawa$^1$, Sergei B. Popov $^{2}$}
\affil{$^1$ Department of Physics, Purdue University, 
 525 Northwestern Avenue,
West Lafayette, IN
47907-2036, USA; lyutikov@purdue.edu
\\
$^2$ Sternberg Astronomical Institute,
Lomonosov Moscow State University, Universitetsky prospekt 13, 119991,
Moscow, Russia}

\begin{abstract} 
We   discuss  possible  association of  fast radio bursts (FRBs) with supergiant pulses 
emitted by young pulsars (ages $\sim$ tens to hundreds of years)   
born with regular magnetic field but 
 very short -- few milliseconds  --  spin periods.
FRBs are 
extra-Galactic events  coming from  distances $d \lesssim 100$~Mpc.   
Most of the dispersion
measure  (DM) comes from the material in the freshly ejected SNR shell;  
for a given burst the  DM should decrease with time.
FRBs are not expected to be seen below $\sim 300 $ MHz due to free-free
absorption in the expanding ejecta.  
A supernova  might have been detected years before the burst;  
FRBs  are mostly associated with star forming galaxies.

The model requires that some pulsars are born with very fast spins, 
of the order of few milliseconds.
The observed distribution of spin-down powers 
$\dot{E}$ in young energetic pulsars is consistent with equal birth rate 
per decade of $\dot{E}$. Accepting this injection spectrum and 
scaling the intrinsic brightness of FRBs  with  $\dot{E}$,  
we predict the following properties of a large sample of FRBs:  
(i) the  brightest observed events come from a broad 
distribution in distances; (ii) for repeating bursts  
 brightness either remains nearly constant 
(if the spin-down time is longer than the age of the pulsar) 
or decreases with time otherwise; in the latter case DM $\propto \dot{E}$.  
\end{abstract}

  \section{Introduction}
 Fast radio bursts (FRBs)
  \citep{2007Sci...318..777L,2012MNRAS.425L..71K,2013Sci...341...53T,2014ApJ...797...70K,2014ApJ...790..101S} 
are recently identified
  mysterious events that are still waiting to be understood.  
 Let us first summarize briefly
  the main observation properties and the key inferences.  (Though some of
  the inferences listed below are based on single FRBs we assume that these
  are common properties. Also, see below a separate paragraph discussing 
the result of \cite{2016Natur.530..453K}.)
  
\begin{itemize}

 \item FRBs are typically milliseconds long events. 
The observed duration is mostly due to scattering broadening during 
propagation \citep{2015arXiv151107746C}; 
intrinsic width is consistent with the $\delta$-function emission in time. 
Still, the observed duration limits can be translated to limits on
the scale of the emission regions as  
$\leq 10^{8}$ cm in size. 
Relativistic bulk motion  with $\Gamma_b$ would modify this estimate by a factor $\Gamma_b^2$, 
but then the corresponding event rates would increase accordingly. 

 \item  The rates are estimated as $\sim 10^3$~--~$10^4$ per day per sky
above 4 mJy per msec \citep[\eg][and references therein]{2016MNRAS.455.2207R}. 
The upper limit is barely comparable to the SN rate up to $z\leq 1$ 
\citep[\eg][]{2009A&A...499..653B}, however, the short duration is inconsistent 
with the SN explosion. The high rate of FRBs also excludes violent events 
like NS-NS mergers \citep[\eg][]{2014A&A...562A.137F}, 
that are expected to have rates at least hundred times lower 
\citep{1991ApJ...380L..17P,2003Natur.426..531B}.  
Also, the observed rate of FRBs 
is, probably, just a lower limit -- 
there can be an even more numerous population of weaker FRBs  
 \citep[as demonstrated  by the repetitive FRB,][]{2016arXiv160300581S}

\item  Dispersion measure (DM) of FRBs is in the range 
few hundreds to few thousands. 
If DM is due to the intergalactic medium this would place FRBs at 
cosmological distances, $z \sim 1$. 
However, DM is likely to come from the local structures near the sources 
\citep{2015Natur.528..523M}. If DM is local, 
the  isotropic distribution of FRBS on the sky 
\citep{2014ApJ...789L..26P,2015MNRAS.451.3278M}  
implies that the  typical distance  $\geq$ few tens Mpc. 

\item  FBRs are repetitive  but non-periodic \citep{2016arXiv160300581S}; 
present-day overall limits on repeatability 
  \citep{2015MNRAS.454..457P}  corresponds to, approximately, 
not more than a burst per day for bright bursts (in terms of observed bursts; 
for highly beamed emission into an angle $1/\Delta \Omega_b$ 
the intrinsic repetitiveness will be 
larger by  $4 \pi /\Delta \Omega_b$). 
In case of weaker bursts the   repetition rate  can be higher  
\citep{2016arXiv160300581S}.
Repetitiveness also  excludes violent events like compact object
mergers. 

\item  Multi-component structures 
\citep{2015arXiv151107746C} hint at rotation at $\sim$ millisecond periods. 
Multi-component structures  can be used as an argument against catastrophic
models like collapse of a NS to a BH, or deconfinement of matter and
formation of a quark star \citep{2015arXiv151107746C}. 

\item  FBRs have relatively flat spectra with index at least flatter
than -3.2 \citep{2016arXiv160102444C}, or even more flat
\citep{2016arXiv160207544R}.  On the other hand, spectra can 
be highly variable, possibly with narrow spectral component 
\citep{2016arXiv160300581S}.

\item FRBs show both circular \citep{2015MNRAS.447..246P} 
and/or linear polarization \citep{2015Natur.528..523M} 
intrinsic to the source. 
In addition, \cite{2015Natur.528..523M} detected intrinsic 
position angle (PA) rotation during the burst, 
possibly consistent with PA swings observed in pulsar 
\citep{1969ApL.....3..225R}.

\end{itemize}

Many of the  above properties stand in sharp contrast to the recently 
 claimed identification of the FRB host with an
 elliptical galaxy at $z=0.45$  \citep[the claimed DM 
would then be cosmological; the
 event would be  consistent with energetic violent events 
like NS-NS merger,][]{2016Natur.530..453K}. 
 \cite{2016arXiv160208434W} (see also ATel \#8752) claimed that very high
 persistent radio emission is inconsistent with the low star-formation rate
 -- the host is probably an AGN.  
Also, small galactic latitude implies lots of
 scattering in the Galaxy; the time scale of few days is typical for 
AGN intra-day
 variability \citep{2001Ap&SS.278...87J}.  From the theoretical point of
 view, the result of \cite{2016Natur.530..453K} -- the implied very bright
 afterglow, -- contradicts the fact that the rate of FRBs is hundreds of
 times higher than of violent merger events like NS-NS mergers.  Also, the
 afterglow implied by \cite{2016Natur.530..453K} needs about $10^{45}$ ergs
 emitted in radio, different by about two orders of magnitude from the 2004
 flare from SGR 1806-20 produced total $4 \times 10^{43}$ ergs
 \citep{Gaensler1806}.  The SGR afterglow lasted longer, with peak flux
 about 200 times higher than Keane's burst (50 mJy versus 250 $\mu$Jy) from
 a distance more than 100 times closer.
  If the results of \cite{2016Natur.530..453K}, 
are confirmed, that would imply two types of FRB progenitors 
(type I - repeating, type II - non-repeating). 
Below we then limit our discussion to the ``type I'' FRBs. 
   
  \section{FRB emission site: \mss\ of \NSs}

 Given the above inferences  let us estimate
parameters at the source. The key unknown is the distance.
 Given that the DM is local \citep{2015Natur.528..523M}, 
it cannot be used as a
distance estimate.  
Yet, isotropy of the observed events argues for a local cosmological origin.  
As a fiducial value we use a typical distance of $d=100$~Mpc 
and take Lorimer burst \citep[][duration $5$ msec, peak flux $S_{30 \rm Jy}$, 
DM=$375$]{2007Sci...318..777L} as a prototypical example.

 The  instantaneous (isotropic-equivalent)  luminosity $L_{FRB}$   is  then 
\be
L_{FRB} =4\pi d^2  (\nu F_\nu)=  3.4 \times 10^{41}  S_{30 \rm Jy} d_{100 Mpc}^2 {\rm erg \, s}^{-1},
\label{LFRB}
\ee 
while the total radiated energy is 
\be
E_{tot} = 4\pi d^2  (\nu S_\nu) \tau= 1.7 \times 10^{39} 
d_{100}^2  \tau_{5\rm msec} \nu_9 S_{30 \rm Jy}\, {\rm erg}
\label{Etot}
\ee
Where $\tau$ is the burst duration (we normalize it to 5 msec) and $\nu$ is the observation frequency (normalized to 1 GHz).
The energy density at the source corresponding to (\ref{Etot})   is
\be
u_{rad} \sim {E_{tot} \over ( c \tau)^3} = 5 \times 10^{14} erg \, cm^{-3}
\label{urad}
\ee
The brightness temperature
\be
T_b \approx {  2 \pi  d^2  S_\nu \over \nu^2 \tau^2}  {\Delta \Omega \over 4\pi} \approx 5 \times 10^{34} \, {\rm K}
\label{Tb}
\ee
clearly implies  a coherent mechanism. 

Radiation mechanism is likely to include particles in \Bf. 
The \Bf\ energy density corresponding to (\ref{urad}) is 
\be
B_{eq}= \sqrt {8 \pi   u_{rad}} =  \sqrt {8 \pi  } {\sqrt{L} \over c^{3/2} \tau} = 10^8
\, {\rm G}.
\label{B2}
\ee

Another requirement for high  \Bf\ in the emission region 
comes from  
the estimate of the wave intensity parameter 
\be
a = {  e E \over m_e c  \om} \approx 10^5 \gg 1
\label{a}
\ee
where $E = \sqrt{L / ( c^3 \tau^2)}$ is the typical \Ef\ 
in the wave at the emission site  \citep{2014ApJ...785L..26L}.
Since the emission is coherent, 
in unmagnetized plasma  the  emitting particles would have  
highly relativistic  \Lf\ $\gamma_\perp \sim a$ and would quickly lose  
energy through various radiative processes 
(\eg synchrotron in case of large momentum perpendicular to the \Bf). 
In highly magnetized plasma (in the limit $\om \ll \om_B$) the   large value of the 
intensity parameter  (\ref{a}) does not necessarily imply high radiative 
losses of emitting particles. 
Instead of oscillation under the influence of the \Ef\ of the wave 
with \Lf\ $\gamma \sim a$, in a high \Bf\ particles experience 
$E\times B$ drift;  we need then  to replace in Eq. (\ref{a}) 
$\om\rightarrow \om_B$; using (\ref{B})  we find 
$a \sim 1/\sqrt{8\pi}$. 
(Also, as is the case for pulsar radio emission, 
the radiation should escape induced Compton scattering in the 
wind \citep{1978MNRAS.185..297W}; this can be achieved  by 
sufficiently fast and/or rarefied  
wind \citep{1992ApJ...395..553S}.

The above estimates, by exclusion, leave only \mss\ of \NSs\ as viable {\it
loci} of the generation of FRBs
\citep{2010vaoa.conf..129P,2015ApJ...807..179P,2016MNRAS.457..232C,2016arXiv160300581S}

\subsection{Two possible mechanisms for FRBs}

\subsubsection{Magnetically and rotationally powered: magnetars versus ``pulsars on steroids''}

Identification of FRBs with \NS\ and evidence against catastrophic events
(collapse, coalescence, etc.) leave two possible mechanisms: 
(i) radio emission accompanying giant flares in magnetars 
\citep{lyutikovradiomagnetar,2010vaoa.conf..129P,2014MNRAS.442L...9L,2012MNRAS.425L..71K,2015ApJ...807..179P}; 
(ii) Giant pulses (GPs) analogues emitted by young pulsars 
\citep{1995ApJ...453..433L,2004ApJ...616..439S,2007A&A...470.1003P}, 
as discussed by \cite{2016MNRAS.457..232C,2016MNRAS.458L..19C}.
Importantly, both scenarios imply repetitiveness and 
FRBs relation to young \NSs\ with particular properties -- 
high \Bfs\ in the case of magnetars, or high spin-down energies 
in the case of GPs. 
These two possibilities rely on different source of energies for FRBs: 
strong magnetic fields in case of magnetars and rotational energy 
in case of GPs. 

Comparing properties of FRBs with the radio emission of magnetars and/or GPs
can, in principle, be used to favor one of the models, magnetically or
rotationally powered, as we discuss next.  
However, lack of understanding of
mechanisms of radio emission from \NSs\ is a major impediment to this
possibility
\citep[\eg][]{1995JApA...16..137M,1999MNRAS.305..338L,1999ApJ...521..351M,2015SSRv..191..207B}. 
Note, that coherent curvature emission by bunches is not considered a viable
emission mechanism \citep{1992RSPTA.341..105M,1999ApJ...521..351M}.

\subsubsection{Different radio emission mechanisms}

Let us briefly outline our current understanding of the radio emission from
\NSs, a long-standing problem in astrophysics.  Qualitatively we can
identify {\it three} types/mechanisms of radio emission in \NSs: (i) normal
pulses, exemplified by Crab precursor  (coming from opened field
lines, probably near the polar cap, having log-normal distribution in
fluxes), see \cite{1996ApJ...468..779M}; (ii) GPs, exemplified by Crab Main
Pulses and Interpulses (coming from outer \ms, near the last closed field
lines; having power-law distribution in fluxes \citep{1995ApJ...453..433L};
possibly with a special subset of supergiant pulses,
see \cite{2012ApJ...760...64M}; sometimes GPs show narrow spectral structure,
see \cite{2007ApJ...670..693H,2007MNRAS.381.1190L}; (iii) radio emission from magnetars
(coming from the region of close field lines, variable on secular times
scales and having very flat spectra), \eg\ \cite{2006Natur.442..892C}.

Though comparison of these general properties of pulsar radio emission with
FRBs is surely inconclusive, we favor the Giant Pulses model for the
following reasons.  (i) Similar time scales of GPs and FRBs, when allowed
for propagation broadening \cite{2015arXiv151107746C}.  (ii) Polarization:
similar to FRBs \citep{2015MNRAS.447..246P,2015Natur.528..523M}, GPs have
strong polarization signals \citep{2007whsn.conf...68S}, often switching
between linear and circular polarization.  (iii) Association of FRBs with
rotationally powered GPs allows testifiable predictions to be made, based on
the possible scaling of the emitted intensity with the spin-down power. 
(iv) non-detection of radio emission during SGR 1806-20 giant flare
\citep{2016arXiv160202188T} provides arguments against the magnetar
association.  \citep[However, note that][searched only for simultaneous
$\gamma-$ and radio signals.  If there is a delay between them, then the
argument is not applicable.  Also, comparison with just one burst of one SGR
can be not very constraining for the whole population.]{2016arXiv160202188T}
(v) Though the inferred flatter spectra of FRBs \citep{2012MNRAS.425L..71K}
make them resemble magnetar radio emission \citep{2006Natur.442..892C},
below we argue that this can be explained by the low frequency free-free
absorption. (vi) Possible narrow spectral features in FBRs \citep{2016arXiv160300581S} resemble  those seen in Crab GPs
\citep{2007ApJ...670..693H}.


  \section{The working  model: giant pulses from young energetic pulsars}

   In the following we  further discuss the  possibility that FRBs are
(super)-giant pulses from energetic newborn pulsars.

 \subsection{DM, RM and free-free absorption  from SNR}

Given that the DM comes from the local environment  
\citep[][]{2015Natur.528..523M} and that
FRBs are related to \NSs, how the values of DM$\sim $ hundreds can be
achieved?   
Both Galactic and intergalactic contributions to the DM 
from the distances of $\leq 100$ Mpc is expected to be typically  $\sim$ tens. 
(High values of DM for some 
Galactic  pulsars \citep{ATNF} is due to our location in the plane of the 
Galaxy and the fact that many pulsars are located within the Galactic plane.)
Typical values of DM of Galactic pulsars is $\sim $ tens or hundreds (but
normally below 375 -- the lowest DM for FRBs).  
Thus, FRBs cannot come from a general
extragalactic pulsar population, but should come from a special sub-class.

A possible   alternative, that we adopt as the main model, is that the DM
comes from a young SNR ejecta, i.e. from a dense shell around a newborn NS.  
Let us estimate the required time scales. 
Let's assume that a recent SN ejecta expelled mass $M_{ej}$.  If the size of
the SN ejecta is $r$, the corresponding DM is
\be
  {\rm DM} \approx  { M_{ej} \over m_p r^2 }
 \label{dm}
 \ee
 So, the larger is the size, the smaller is the DM. 
For a given DM the size is
 \be
 r= \sqrt{ M_{ej}/m_p} {1 \over \sqrt{ {\rm DM}}} = 0.34 {\rm pc} \, \sqrt{ m_\odot}
{\rm DM}_{375}^{-1/2},
 \label{rmax}
 \ee
 where ${\rm DM}_{375} = {\rm DM}/375$ and $m_\odot= M_{ej}/M_{\odot}$.

 Swept-up mass
 \be
 {  M_{swept} \over M_{ej}} = \sqrt{ M_{ej}/m_p} {  n_{ISM} \over 
{\rm DM}^{3/2} {\rm pc}^{3/2}} = 4.5 \times 10^{-4} n_{ISM} \sqrt{ m_\odot} \ll 1,
 \ee
 where $n_{ISM} $ is the circumburst ISM number density.
 So the motion is typically ballistic with velocity
 \be
 v_{ej} = \sqrt{ 2 E_{ej} \over M_{ej}}.
 \ee
 To reach the size (\ref{rmax})
 it takes
 \be
 t= {M_{ej} \over \sqrt{ 2 {\rm DM} E_{ej} m_p}} = 35 {\rm yrs}\,m_\odot
 \label{t}
 \ee
(for $E_{ej}  =10^{51} $ erg.)

\cite{2015Natur.528..523M}  claimed  ${\rm RM}= 180$ rad/m$^2$ and 
${\rm DM} =600$ pc cm$^{-3}$ in the circumburst surrounding; this     
implies the average \Bf\
  \be
  B = 2 \pi {m_e^2 c^4 \over e^3} {{\rm RM} \over {\rm DM}} = 
3 \times 10^{-7} {\rm G},
  \label{B1}
  \ee
  below the typical Galactic field of $\mu$G. 
One possible explanation is that  the  \Bf\ that produces the RM is 
confined to the expanding envelope -- it is  then expected to be in 
toroidal direction, perpendicular to the line of sight.

The free-free optical depth through an expanding SN shell  is sufficiently 
small at  $\sim$ GHz frequencies \citep[][Eq. 1.223]{1999acfp.book.....L}
 \be
 \tau =   8 \times 10^{-2}  n^2 \nu ^{-2.1} r  T ^{-1.35} = 0.05 \,  {\rm
DM}_{375} ^{5/2} m_\odot^{-1/2}
\nu_9^{-2.1}.
 \ee  
 Note that the free-free optical depth  becomes of the order of unity 
at frequencies  $\leq 300 $ MHz. 
This might explain the fact that low frequency  observatories like LOFAR  
and MWA did not see FRBs \citep{2015MNRAS.452.1254K,2016arXiv160207544R}, 
and  that the lowest frequency of FRB detection so far  
is $700$MHz \citep{2015Natur.528..523M,2016arXiv160207292C}.
The plasma frequency in the ejecta is 
 $\om_p = \sqrt{4 \pi n e^2/m_e} = 10^6 {\rm DM}_{375}^{3/4} ({
M_{ej}/M_{\odot}})^{-1/4}$ rad s$^{-1}$, 
so the source is transparent to radio waves at $\sim 1 $ GHz.

We conclude that very young SNRs, at ages  tens to hundreds of  years 
can provide the DM of the order of the observed values.

 \subsection{Pulsar physics}
 
Above  we have  established that the observed properties of FRBs are
  consistent with SN environment $\sim $ tens of years after the explosions. 
  Next, let us discuss how pulsar physics fits with these estimates.  We
  hypothesize that FRBs are rare (super)giant pulses-like events whose
  luminosity
$L_{FRB} $  scales with the spin-down power of a pulsar, 
$L_{FRB} = \eta \dot{E}$, $\eta \ll 1$ 
\citep[we note that this is not the case 
for the bulk of the pulsar population][]{ATNF}.
 
 For Crab pulsar the peak GP fluxes $S_\nu$  exceed Mega-Janskys 
\citep{2003Natur.422..141H,2007whsn.conf...68S,2012ApJ...760...64M}. 
The corresponding {\it instantaneous}  efficiency
 \be
 \eta = {L_{GP} \over \dot{E}_{\rm Crab}} =
{ \nu c^3 d_{\rm Crab}^2 S_\nu P_{NS}^4 \over 4 \pi^3 B_{NS}^2 R_{NS}^6} \approx
10^{-2},
 \label{eta1}
 \ee
 where subscript NS refers to the \Bf\ on the surface, 
radius and period of Crab pulsar.
 Since the GP duration is much smaller than the period, 
the average efficiency is much smaller than (\ref{eta1}) 
by $\Delta \Omega /4 \pi = (\Delta \theta)^2/4 \approx $ few $\times 10^{-7}$
where $\Delta \theta \approx  2 \pi \tau_{GP}/P_{NS} \approx 10^{-3}$ is the relative 
active  phase of the GP, $\tau_{GP} \sim $ few $\mu$sec is the GP duration.
 Thus, instantaneously, pulsars GP luminosity can reach $\eta \sim $ few
percent of the spin down power.  

By analogy with Crab GPs we expect that the
intrinsic FRB duration is smaller than the \NS\ spin.  (FRB duration
is consistent with $\delta$-function pulse smeared by propagation effects, 
\citealt{2015arXiv151107746C}.)
Normalizing the FRB duration to the \NS\ spin, the required \Bf\ is
 \be
 L_{FRB} = \eta \dot{E} \rightarrow B_{NS} = {c^{3/2} d \sqrt{(\nu F_\nu)} P_{NS}^2 \over  2 \pi^{3/2} R_{NS}^{3/2} \sqrt{\eta}} =
 2  \times 10^{13}\, d_{100 \rm Mpc} F_{30 \rm Jy}^{1/2} \tau_{5 \rm msec} ^2 \sqrt{\nu_9} \eta_{-2} ^{-1/2} \, {\rm
 G}.
 \label{B}
 \ee
 The \Bf\ (\ref{B}) is somewhat larger than the typical $\sim 10^{12}$ G of young pulsars 
(recall that this estimate uses the 
FRB duration as an estimate of the spin period, $B \propto   \tau^2$), yet  
it is well within the overall distribution of rotationally-powered pulsars, 
especially given the uncertainties on other parameters. 
Also, if intrinsic duration of the FRB is much smaller than the period, 
the estimate of the \Bf\ (\ref{B}) decreases accordingly. 
  The corresponding spin-down time is  
  \be
 \tau_{SD} = { \pi \eta I_{NS} \over d^2 F_\nu \mu P^2} \sim  {\rm few \,
years}.
 \ee
 (The most important constraint in the above estimate comes 
from equating FRB duration 
with the rotation period of a \NS. 
Duration of  GPs is typically much shorter that the period.)
Thus, {\it if FRBs are powered by the rotation of a \NS\ it is required that 
(some fraction) 
of pulsars is born with millisecond periods (and normal \Bfs)}.
 
Observationally,  the initial periods of \NSs\ are generally unknown  \cite[the 
fastest young pulsar PSR J0537-6910 has 16 msec
spin,][]{1998ApJ...509L.109W}. 
The main way to probe initial spin periods of NSs is to obtain an
independent age estimate (a SNR age, or a kinematic age, etc.) and to apply a
usual magneto-dipole formula (with braking index three). The largest set of
such calculations for NSs in SNRs has been presented by
\cite{2012Ap&SS.341..457P}. Despite, on average objects in this study
appeared to have initial spin periods $\sim 0.1$~s, significant fraction of
analyzed sources can have very short initial periods.  
Even if pulsars are born with millisecond periods 
(but ``regular'' \Bfs\ of $\sim 10^{12}$ G) they are
 expected to  quickly
(within tens of years) spin down to periods larger than 
$\sim 10$ msecs 
\citep{1993MNRAS.263..403L,2001ApJ...549.1111L,2008AdSpR..41..503V}. 
\cite{1999A&A...346L..49A} did argue in favor of fast initial spin in Crab. 
Theoretically, 
some simulations do predict rotation of \NSs\ with the spins in the
millisecond range \citep[\eg][]{2016arXiv160102945C}.
 
 We conclude that a population of young pulsars  (ages tens to hundreds of
years) with \Bfs\ typical of the observed population of young pulsars, but
with spin periods in the few millisecond range is a viable source of FRBs
\citep[see also][]{2016MNRAS.457..232C,2016MNRAS.458L..19C}.
  
 \subsection{Frequency of occurrence}

Let us do an estimate of the frequency of occurrence assuming  
that FRBs come from  young powerful \NSs\ in the local universe,   
from distances  $ d \lesssim 100 $ Mpc.  

\cite{2012ApJ...757...70D}
estimate  the core-collapse SN rate $\sim 3 \times
10^{-4}$~yr$^{-1}$~Mpc$^{-3}$. Then in 100 Mpc we expect $\sim300$ SN per
year (or one per day). If we assume that all young PSRs can produce strong
bursts up to the age $\sim 30$~yrs, 
then we have $\sim10^4$ such sources inside 100 Mpc.
To have a rate of few$\times 10^3$ FRBs per day, 
each pulsar needs to produce one-two
bursts per day, roughly consistent with current overall limits
\citep{2015MNRAS.454..457P}. If we slightly increase the limiting distance
(say, up to 200 Mpc),
then we can obtain a more comfortable fraction ($\sim0.1$)
of young PSRs having large $\dot E$, and so producing strong bursts.


One can  estimate the rate of supergiant pulses from a given pulsar 
following the data given by \cite{2016MNRAS.457..232C}.
The Crab pulsar produces GP with flux $\sim100$~--~$200$~kJy once per
hour (the brightest one is slightly than an order of magnitude more
luminous). 
If we take a pulsar with the same field but spin period 
$\lesssim 2$~msec,
then it has $\dot E$ $\sim 10^5$ times larger. So, we can expect from the same
distance (2 kpc) flux $(1$~-~$2)\times10^{10}$~Jy. 
If we now consider distances 100-200 Mpc, then the flux is about few Jy.
And the brightest --- about few tens of Jy. Well in the range of FRB fluxes. 
Flux distribution for the Crab pulsar is roughly $\propto
S^{-3}$\citep{2016MNRAS.457..232C}. I.e.,
much brighter bursts (for example, like the Lorimer burst) might be rare --
once in several months from the same source. 

 In the case of the 
repeating FRB 121102 \citep{2016arXiv160300581S}, 
the  observed rate $\sim 3$ hr$^{-1}$ is high, but potentially consistent
with expectations for young PSRs since most bursts are of lower intensity,
and some sources can be more active than average. 

\section{Expected statistical properties of FRBs} 

Currently, only a handful of FRBs is known 
\citep{2016arXiv160103547P}.  
Let us now calculate statistical properties of FBRs expected in our model,
 which can be later tested with larger statistics.  
The key assumptions here is that
the intrinsic luminosity of an FRB is proportional to the spin-down power
$\dot{E}$.

    \subsection{Injected  and observed distribution  in spin-down energy $f(\dot{E})$}

Let's assume that pulsars are produced with a rate 
$f_{inj}(\dot{E})$ (per unit time, per unit volume, 
per unit range of $d\dot{E}$). 
The Boltzmann equation for the evolution of the number 
density $f(\dot{E})$ reads
\be
\partial_t f + \partial_{\dot{E}} \left( ( \partial_t \dot{E} ) f\right) =
f_{inj}.
\ee
Assuming constant \Bf\ (since we are interested in very young pulsars 
we neglect possible \Bf\ decay), the spindown power evolves according to 
\be 
 \partial_t \dot{E}  = - {  4 B_{NS}  R_{NS}^3 \over c^{3/2} I_{NS}}  \dot{E}^{3/2} 
 \ee
 (the subscript $NS$ refers to the surface properties of the \NS). 
 
The steady state kinetic equation in $\dot{E}$ coordinates,
\be
\partial _{\dot{E}} \left( \partial _t (\dot{E}) f(\dot{E})  \right) = f_{inj}(\dot{E}),
\ee
with the injection spectrum 
\be
f_{inj}(\dot{E})  \propto \dot{E}^{-\beta}
\ee
has a solution
\ba && 
 f(\dot{E})  \propto c_1  \dot{E}^{-1/2 -\beta} + c_2  \dot{E}^{-3/2 }, \beta \neq 1
  \label{1}
  \\
  && 
  f(\dot{E})  \propto {\ln (\dot{E}_0/\dot{E}) \over \dot{E}^{3/2}}, \beta =1
  \label{2}
 \ea
 for the $\beta \neq 1$ case the term with $c_2$ 
is from the solution of the homogeneous equation;  
for the $\beta = 1$ case $\dot{E}_0$ is an integration constant.
 

  \subsection{Observed distribution in $\dot{E}$: the inferred injection spectrum}
\label{Observationa}


 As we  discussed above,  at high values of $\dot{E}$ the observed and the
injection spectra of pulsars are related by Eqs. (\ref{1}-\ref{2}).  The special
case of $\beta=1$ (same number of pulsars born per decade of $\dot{E}$)
is particularly interesting.  Next, we demonstrate that the observed
distribution of high spin-down power pulsars is indeed consistent with such 
fairly flat injection spectrum.

\begin{figure}[htbp]
\begin{center}
\includegraphics[width=0.49\linewidth]{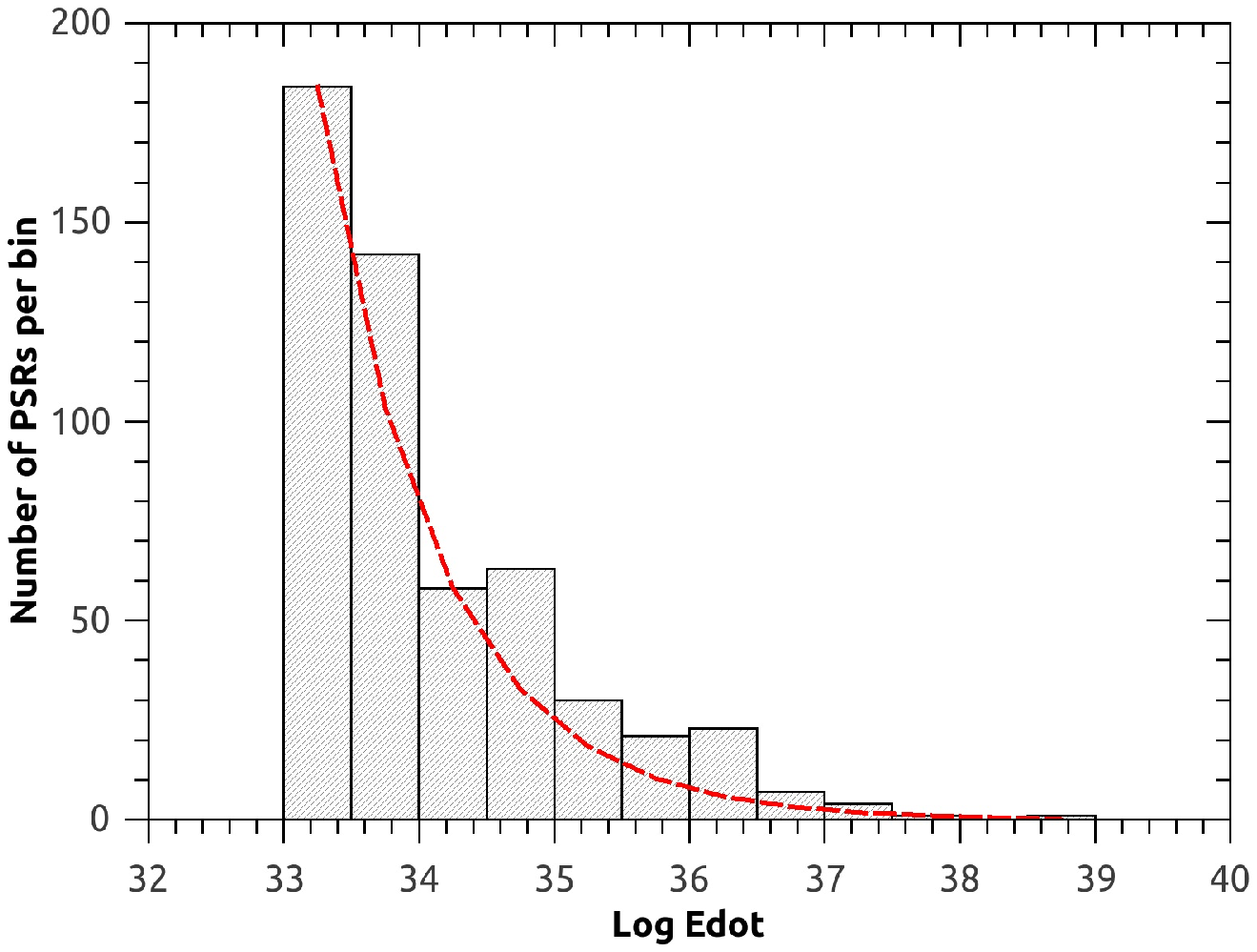}
\includegraphics[width=0.49\linewidth]{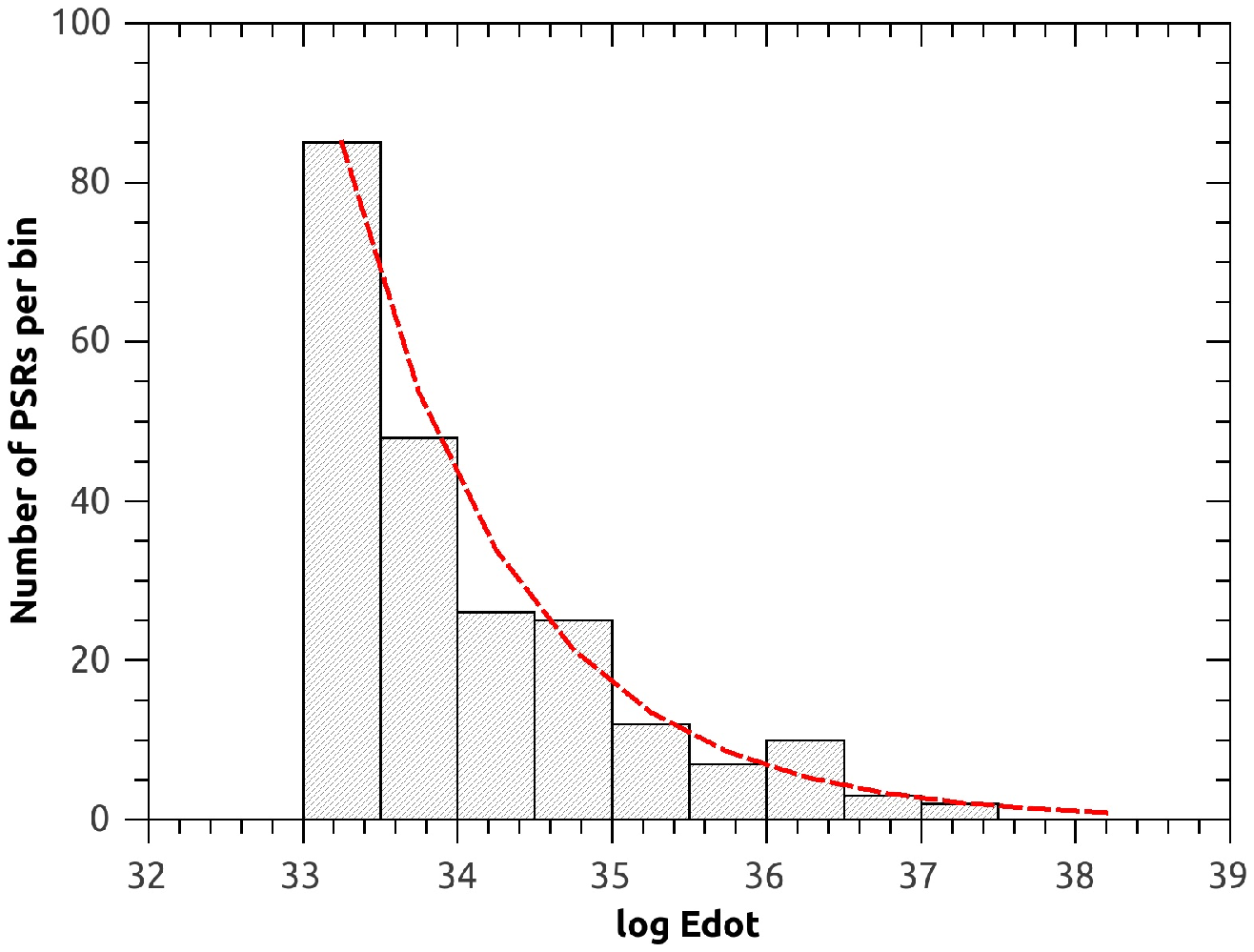}
\caption{ {\it Left Panel}. Differential $\dot E$ distribution above $\dotE
=10^{33}$~erg~s$^{-1}$ in log-scale. Selected PSRs have $B>10^{11}$~G,
$S_{1400}>0.1$~Jy. Dashed line corresponds to the law $dN/d(\dot
E) \propto \dot E^{-1.5}$.
{\it Right Panel}. Differential $\dot E$ distribution above $\dotE
=10^{33}$~erg~s$^{-1}$ in log-scale. Selected PSRs have $B>10^{11}$~G,
$S_{1400}>0.1$~Jy, and distances $>7$~kpc. Dashed line corresponds to
the law $dN/d(\dot E) \propto \dot {E}^{-1.4}$.}
\label{edot}
\end{center}
\end{figure}

We use the ATNF catalogue \citep{ATNF} to obtain the $\dot E$ distribution.
Since we are interested in young powerful pulsars, we use the high-energy tail
of the distribution. We made several different radio pulsar samples from 
the ATNF catalogue to study
the $\dot E$ distribution. Two of them are shown in Fig. \ref{edot}.
Number distributions of pulsars per logarithmic bin is typically
roughly $\propto \log \dotE ^{-0.5}$ which corresponds to
 $dN/d(\dot E)\propto\dot E^{-1.5}$ (red dashed curve 
in left panel of Fig. \ref{edot}). 
For some samples, for example the sample of PSRs with $B>10^{11}$~G,
$S_{1400}>0.1$~Jy and distances $>7$~kpc (Fig. \ref{edot}, right panel)
a better fit is $dN/d(\dot E)\propto\dot E^{-1.4}$, 
still very close to the -3/2 law.

In addition, we analyzed period distribution of short period pulsars.
Observed distribution in $P$ for short period pulsar (from $\sim0.03$ to
$\sim 0.2$~s)
is $f(P) \propto P^{1/2}$. 
Though this range of periods is also populated by older objects, it
contains many young sources, in correspondence with estimates by
\cite{2012Ap&SS.341..457P}. Thus, it can be used to estimate the initial
$\dot E$ distribution. 

Since for the simple magneto-dipole formula
$P \propto \dot{E} ^{-1/4}$ this translates to 
\be
f(\dot{E}) \propto \dot{E}^{-11/8},
\ee
This is  sufficiently close to the $\alpha=-3/2$ law. 
Thus,   we conclude that the observed distribution of fast pulsars 
is consistent with injection parameters 
$\beta =1$, $f_{inj} \propto 1/\dot{E} $ 
(equal number of newborn sources per decade of  $\dot{E} $).  The distribution in spin-down power  $f(\dot{E})$ can be related to the multivariate distribution in period and \Bf\ $f(B,P)$,
 see \S \ref{fBP}.


 \subsection{Homogeneous source distribution}

\begin{figure}[h!]
\begin{center}
\includegraphics[width=0.45\linewidth]{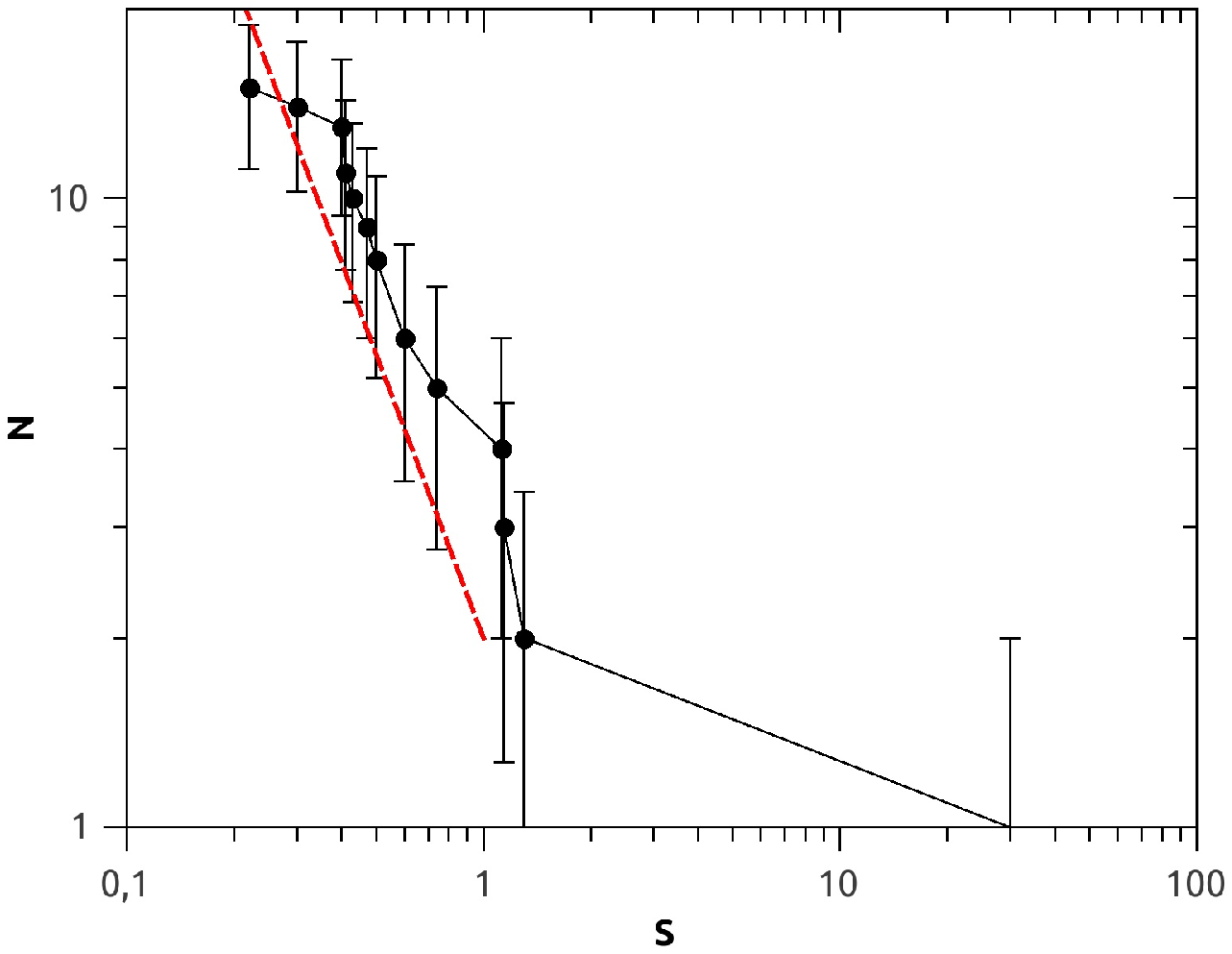}
\includegraphics[width=0.45\linewidth]{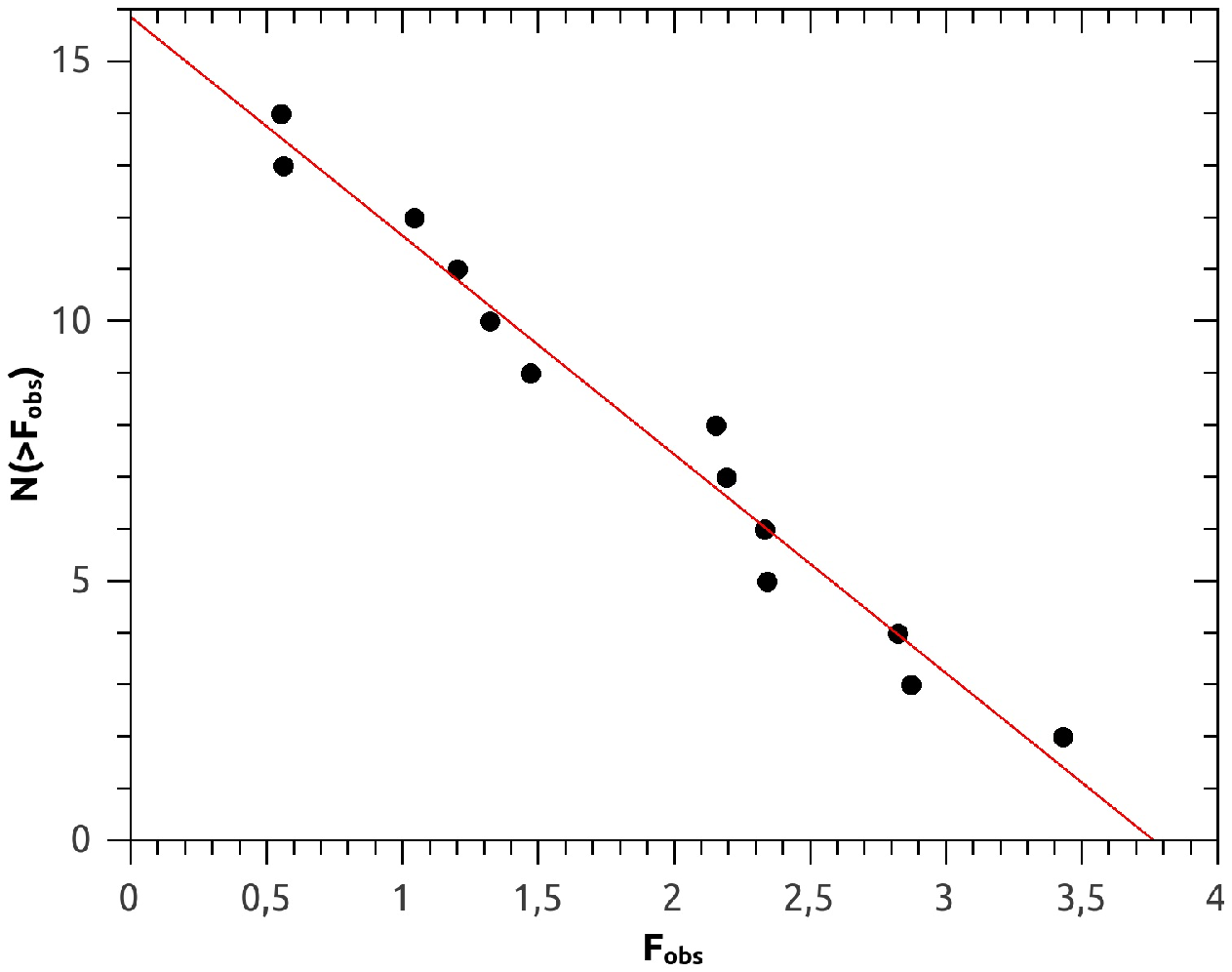}
\caption{{\it Left Panel}. 
Log $N$ -- Log $S_\mathrm{peak}$ distribution for FRBs.
The observed distribution is consistent with homogeneous distribution of
sources. The  brightest  burst, the Lorimer  burst, is excluded from  the fit. {\it Right Panel}. 
 Distribution $N(>F_\mathrm{obs})$--$F_\mathrm{obs}$ on the linear-linear scale. Two brightest sources are not included.
}
\label{lnls}
\end{center}
\end{figure}

We analyzed the observed Log $N$ --- Log $S_\mathrm{peak}$ distribution of FRBs, Fig.
\ref{lnls}, using  the  FRB catalogue \citep{2016arXiv160103547P}.
If we exclude the brightest  burst, the Lorimer  burst,
 the distribution  is compatible with the
isotropic $-3/2$ law, Fig. \ref{lnls}. In addition, the distribution in $F_{obs}$ for 13 dimmest sources has a very peculiar form: it is linear in the linear scale. However, statistics is low. 
Deviations from the 3/2 law are expected in  radio surveys, 
since  the effective area of the telescope beam  is a strong function of
flux - super-bright sources (like the Lorimer burst) 
can be found further away from  
 the centre of the radio beam. This biases the Log $N$ -- Log $S$ 
towards flatter apparent spectral distributions  
\citep[][also found flatter distribution]{2016arXiv160206099L}. 
In addition, low statistics seems to bias  
the Log $N$ -- Log $S$ distribution towards flatter indices.  
We have performed a number of trials selecting 16
sources from various luminosity distributions and fitting with the
power-law.  We notice that, first, for the small number of sources the
average value of the power-law index was below $3/2$ and, second, the
standard deviation for 16 sources was $\sigma \approx 0.2$.  
We
conclude that the observed distribution is consistent with homogeneous
source distribution.
 
  \subsection{DM-peak flux correlation}
Combining expressions for DM (\ref{dm}) 
with spin-down power ($\dot{E}_0$ is the value at birth, 
$\tau$ is the initial spin-down time) 
\be
\dot{E} = {\dot{E}_0 \over (1+t/\tau)^2},
\ee
we find
\be
{\rm DM}=  {M_{ej} ^2 \over 2 E_{ej} m_p \tau} { \dot{E} \over (\dot{E}+\dot{E}_0)^2} 
\ee
(recall that we use $\dot{E}$ as a proxy for peak luminosity). 
Thus, for times $t\ll \tau$, when $\dot{E} \approx \dot{E}_0$ we expect 
that DM is independent of the $\dot{E}$ and, under assumptions of the model, 
of $S_{peak}$. For longer times, DM should decrease with
$\dot{E} $ (and $S_{peak}$),  DM $\propto \dot{E}$. 


\begin{figure}[htbp]
\begin{center}
\includegraphics[width=0.69\linewidth]{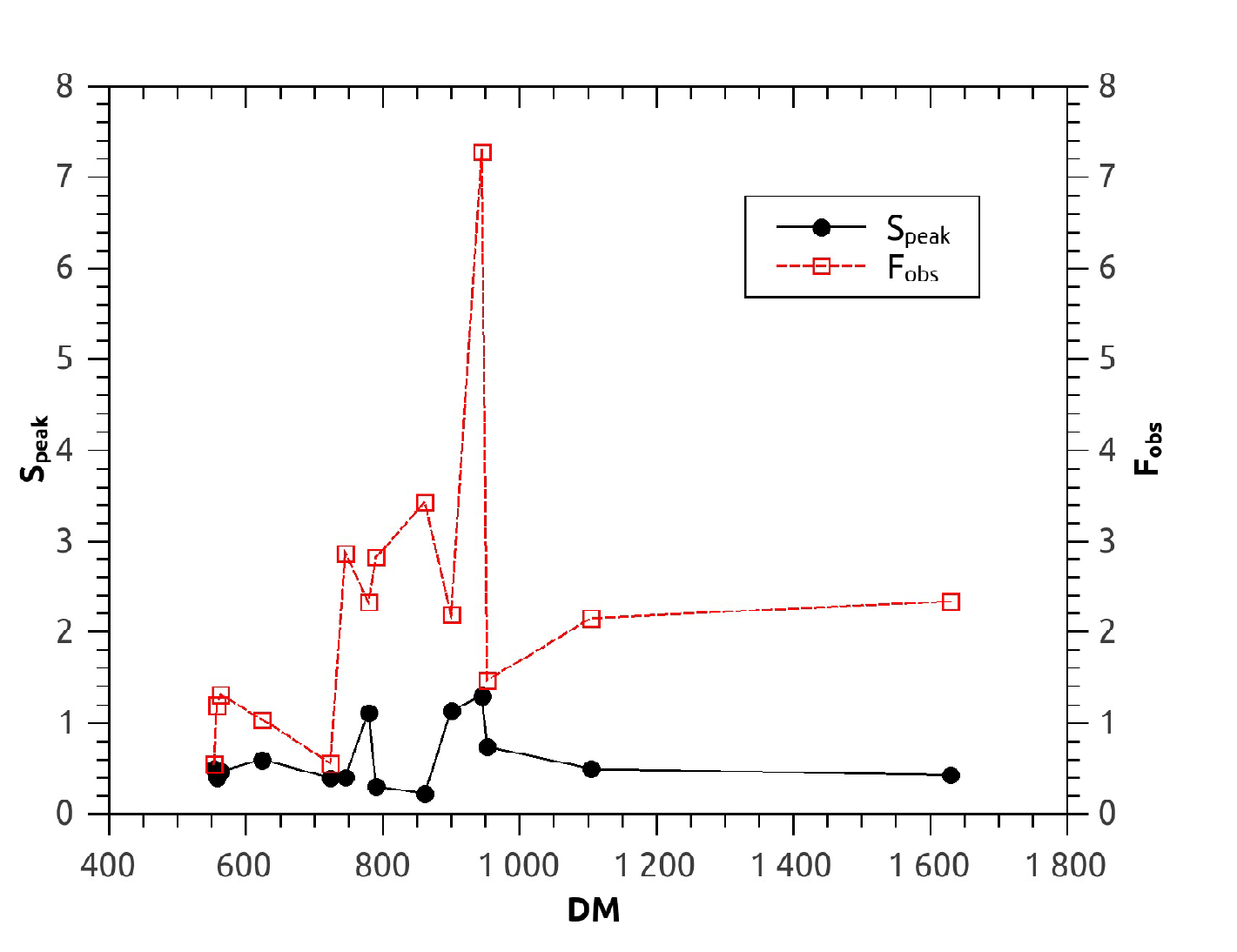}
\caption{
Peak luminosity, $S_{peak}$, and fluence, $F_{obs}$ 
vs. dispersion measure (from the FRB
catalogue Petroff et al. 2016). The Lorimer burst (FRB 010724) is removed from
the plot as it is too bright in comparison with the others.}
\label{dm_sf}
\end{center}
\end{figure}
In Fig. \ref{dm_sf} we plot the observed  peak luminosity, $S_\mathrm{peak}$, 
and fluence,
$F_\mathrm{obs}$, vs. dispersion measure. Obviously, there is no strong
dependence of $S_\mathrm{peak}$ and $F_\mathrm{obs}$ on DM. 
By itself, this behavior of DM excludes  models in which
dispersion measure is a proxy of distance (as dispersion happens in the
extragalactic medium) and bursts are more or less standard candles.  

  



 \subsection{Logarithmic injection in $\dot{E}$ ($\beta=1$):  implications for the radial distribution of brightest sources}

The spin-down power 
distribution $f(\dot{E}) \propto \dot{E}^{-3/2}$ is, in many
respects, a special case: the dipole spin-down law (with constant \Bf)
singles out this solution as a special one (this is a solution of a
homogeneous Boltzmann equation for the pulsar flow); also, it is
consistent with the observed spin-down distribution of fast pulsars (see 
\S \ref{Observationa}). The observed spectrum $3/2$ implies, approximately, a
special injection spectrum $\beta =1$ -- equal number of newborn pulsars per
decade of $\dot{E}$.  Next we calculate the expected observed properties of
FBRs for this special injection spectrum $\beta =1$ (again, assuming that
intrinsic brightness correlates with spin-down luminosity).

First, we evaluate the expected distribution of distances for a given
observed flux.  In an unlimited volume the distribution of fluxes follows
the $N(>S) \propto S^{-3/2}$ law independent of the intrinsic luminosity
distribution.  On the other hand, the distribution of sources contributing a
given flux in distances depends on the intrinsic luminosity function.
 The injection spectrum  $f_{inj} \propto 1/\dot{E} $  
translates into steady state observed distribution (\ref{2}) 
$  f(\dot{E})  \propto {\ln (\dot{E}_0/\dot{E}) / \dot{E}^{3/2}}$.
 Neglecting for a moment slowly varying logarithm, the case $\alpha=3/2$ 
turns out to be an  interesting special case: the distance  to the nearest  
source  is $r \sim f(\dot{E})^{-1/3}$ and  the 
 observed flux (again, assuming that FRB luminosity follows $\dot{E}$)
\be
S \propto { \dot{E} \over r^2} \propto { \dot{E}\,  f(\dot{E})^{2/3}}
\ee
For  $f(\dot{E}) \propto \dot{E}^{-\alpha}$  this implies
\be
S \propto \dot{E}^{(1-2\alpha/3)} \propto r^{-2 +3/\alpha}
\ee
So, for $\alpha< 3/2$,  $S$ increases with $r$ -- the brightest sources are far
away.  For the special case $\alpha = 3/2$, the observed brightness is
independent of the distances, $S \propto r^0 \propto \dot{E}^0$ (in a larger
volume there are brighter sources -- this a variant of a so-called Malmqvist
bias).

Thus, we expect that  for the logarithmic injection spectrum, $\beta=1$, at
a given flux the observed sources are distributed over a wide range of
distances.  On a more subtle point, for the logarithmic injection spectrum
the expected $\dot{E}$ distribution (\ref{2}) differs slightly from $3/2$,
by a logarithm; thus we still expect that closer sources are brighter, yet
the brightest sources are distributed over a large distance.  This is
confirmed by our Monte Carlo simulations which we discuss next.

We conclude that for the injection spectrum $f_{inj} \propto 1/\dot{E} $ the
observed brightest sources have very broad spacial distribution - the
brightest one hundred sources (out of approximately a million in the total
sample) are located within $\sim $ a quarter of the test volume.  This is
important: isotropy of FRBs imply that the brightest ones cannot come from
nearby sources
--- the local Universe  is highly inhomogeneous on scales of tens of Mpc.

 \subsection{Monte-Carlo simulations of pulsars'  spin-down and observed brightness distribution}

\begin{figure}[htbp]
\begin{center}
\includegraphics[width=0.45\linewidth]{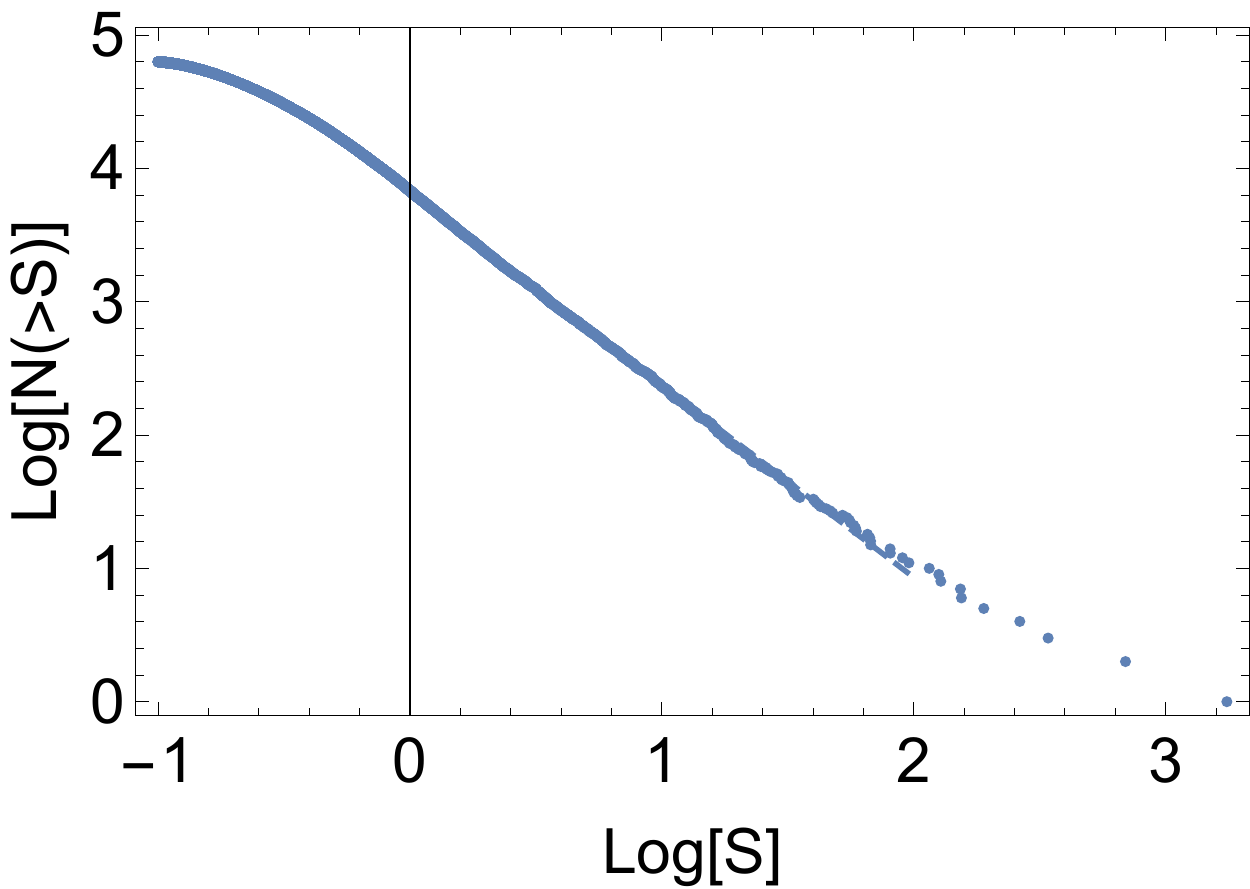}
\includegraphics[width=0.54\linewidth]{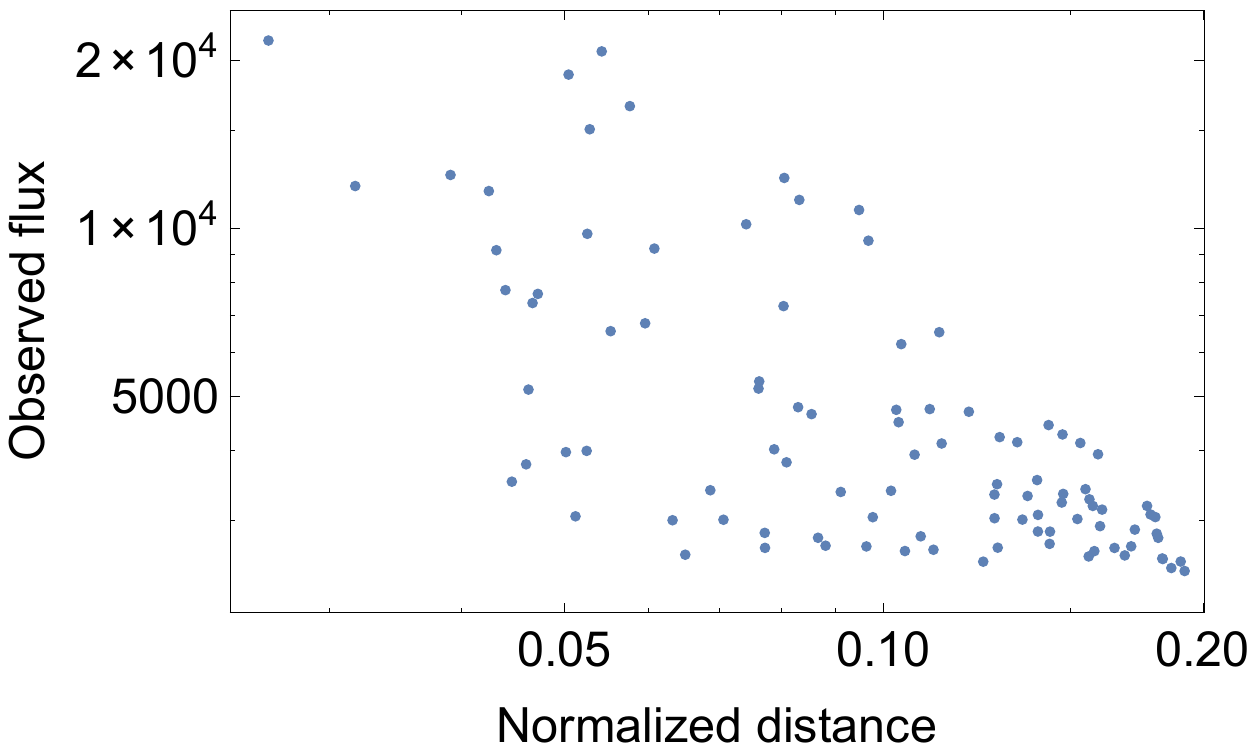}
\caption{ {\it Left Panel}. Observed distribution of fluxes 
$\propto \dot{E}/r^2$ for  injection spectrum (\ref{2}). 
The brightest source are well fitted with $-3/2$ spectrum 
(dashed line, fitted slope $1.53$). 
{\it  Right panel}. Radial distribution of the hundred brightest sources 
(out of total number of approximately one million, 
located at normalized distances between 0 and 1). 
This plot shows that the  brightest sources are, generally, 
located  in a broad range of distances.}
\label{NofSMC}
\end{center}
\end{figure}

To test the spatial distribution of the brightest sources we have conducted
  simulations of pulsar population.  First, to test the spacial distribution of brightest sources  we injected 
pulsars with the expected steady state distribution (\ref{2}) over a range of distances.
The results are shown in Fig. \ref{NofSMC}. Importantly, this confirms that brightest FRBs come from a wide range of distances.

\begin{figure}[htbp]
\begin{center}
\includegraphics[width=0.49\linewidth]{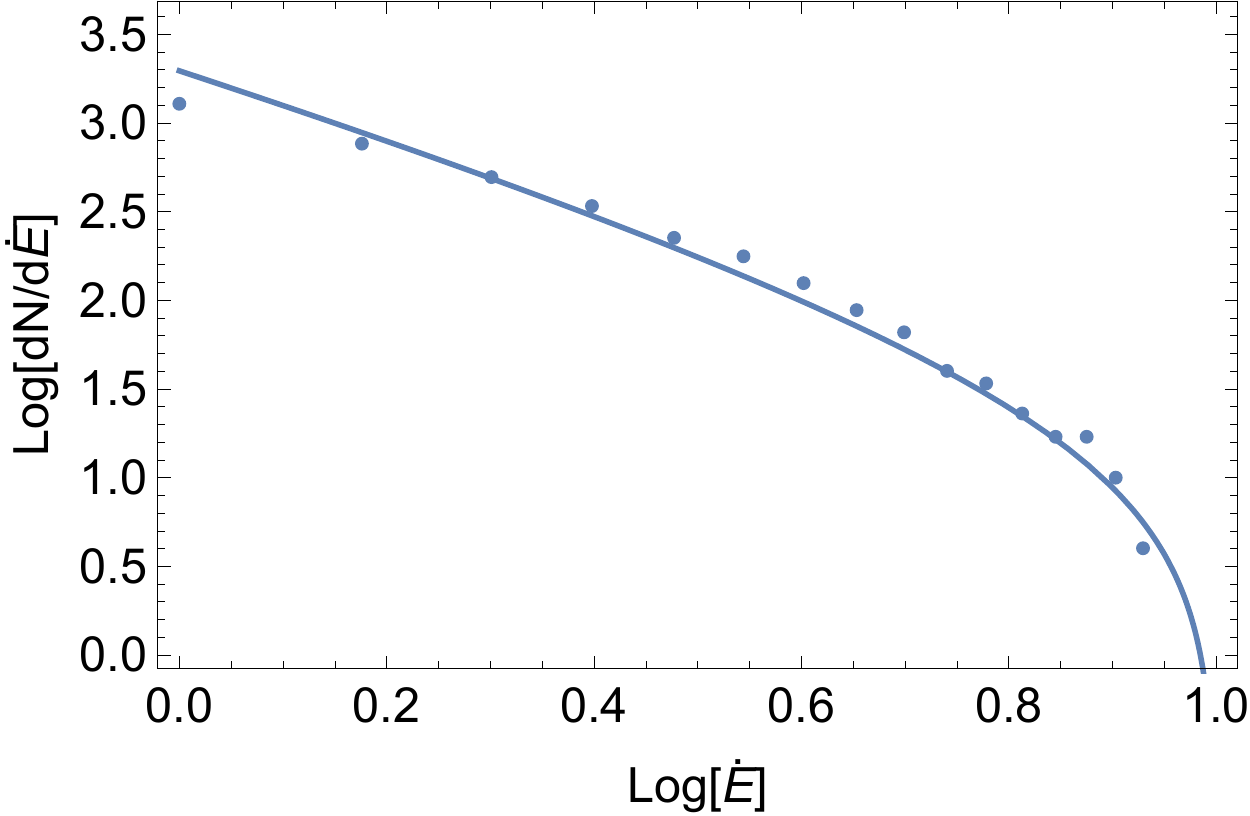}
\includegraphics[width=0.49\linewidth]{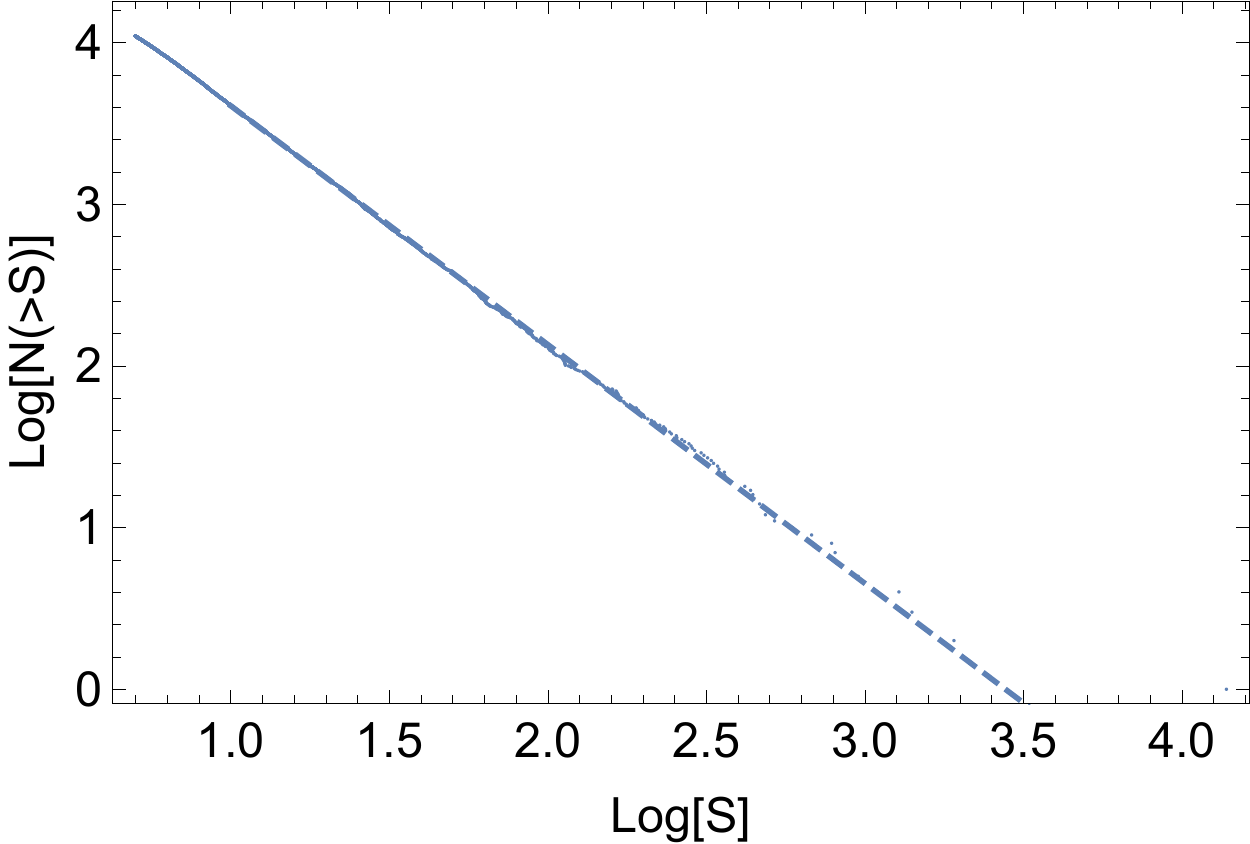}
\caption{ {\it Left Panel}. 
Comparison of MC calculations with spin-down and the expected 
analytical distribution (\ref{2}). Slight disagreement is probably 
due to ``edge effects'' (small dynamical range).  
{\it  Right panel}. Observed distribution of fluxes 
$\propto \dot{E}/r^2$ for MC simulations. 
The high-S tail is fit with power-law $1.47$. }
\label{NofSMC1}
\end{center}
\end{figure}

Second, we did Monte Carlo runs   injecting  \NSs\ by
supernovae and following subsequent spin down evolution.  At each time step we redistribute 
homogeneously a number of \NSs\ in a volume $0<r<1$ with initial
distribution $f_{inj} \propto \dot{E}^{-1}$, $0.1< \dot{E}<1$.  Neutron
stars spin down according to the magneto-dipole 
formula $\partial_t \dot{E} \propto
- \dot{E} ^{-3/2}$.  The observed FRB flux is parametrized with spin-down
luminosity, $S \propto \dot{E}/r^2$.  Sources with flux below some minimal
value are discarded.
 After sufficiently large number of time steps the total number of pulsars 
reaches statistical equilibrium. 
The distribution function $f(\dot{E})$ approaches 
the limit  (\ref{2}), Fig. \ref{NofSMC1}, left panel, while the distribution 
of brightness approaches $- 3/2$ power law, Fig. \ref{NofSMC1}, right panel

In conclusion, our MC simulations confirm the analytical estimates: the
spindown distribution follows (\ref{2}), the brightest observed sources are
distributed over a wide range of distances, and, naturally, that the expected
sources count follows $N(>S) \propto S^{-3/2}$.

 \section{Predictions}
 
 In this paper we argued that  the physical constraints imposed by the
properties of FRBs limit their origin to the \mss\ of \NSs.  Two special
types could satisfy those constraints: fast rotating young \NSs\ (using the
rotational energy to generate FRBs), or very high \Bfs\ \NSs\ --- magnetars
(using the magnetic energy). The key distinction between the two possibilities
would be a detection of high energy emission contemporaneous with an FRB ---
Crab giant pulses do not show high energy signals
\citep{2012ApJ...749...24B,2012ApJ...760...64M,2012ApJ...760..136A}.
 
  In this paper we discussed possible observational features of the  GP-FRB
association.  (i) Since we associate FRBs with recent core-collapse
explosions, we expect that a SN might have been detected years before the
burst.  We encourage observers to search in archives for such correlations.  
(A possible exception to this could be alternative channels of \NS formation, 
like
accretion-induced collapse \citep{1991ApJ...367L..19N}; such events should
have low DMs.  Magnetar activity  is expected to be delayed from  the formation of
a \NS\ by corresponding Hall time, which can vary from years to millennia
depending on the location within the crust \citep{2015MNRAS.447.1407L}.) (ii)
As we expect that distances are $\lesssim 100$~--~$200$~Mpc, then it might
be possible to identify the host galaxy, which will have significant
star formation rate.  (iii) We expect the repetition rate of 
FRBs of
the order of one per day per source.  (iv) For a given FRB source the DM
through a newly ejected SNR should decrease with time (the repeating FRB
121102 did not show such a predicted behavior --- possible mitigating
factors could be: large Galactic contribution, at least 30\% but possibly
higher; older SN, few hundred years; possibly large contribution from the
host galaxy (as opposed to the surrounding SNR).  At the same time the
observed brightness might either be independent of time and of the DM (if
the observation time after a SN is shorter than the initial spin-down time),
or to decrease with time (if the observation time after a SN is longer than
the initial spin-down time).  (v) True distances to FRB sources will show
large variations (not necessarily the closer -- the brighter).  (vi) Some
pulsar are born with very fast spins, of the order of few milliseconds. 
Most of the above predictions assume scaling of intrinsic luminosity with
the spin-down power.

\acknowledgements
ML would like to thank organizers and participants of the workshop 
Transient Bormio 16,   in particular Matthew Bailes, Jason Hessels and   Antonia Rowlinson.  
This work was supported by  NASA grant NNX12AF92G and NSF grant AST-1306672.
  SP thanks profs. K.A. Postnov and V.S. Beskin for discussions. SP was supported by the
Russian Science Foundation,  project 14-12-00146. 
   
 \bibliographystyle{apj} 

\begin{thebibliography}{67}
\expandafter\ifx\csname natexlab\endcsname\relax\def\natexlab#1{#1}\fi

\bibitem[{{Aliu} {et~al.}(2012){Aliu}, {Archambault}, {Arlen}, {Aune},
  {Beilicke}, {Benbow}, {Bouvier}, {Buckley}, {Bugaev}, {Byrum}, {Cesarini},
  {Ciupik}, {Collins-Hughes}, {Connolly}, {Cui}, {Dickherber}, {Duke}, {Dumm},
  {Falcone}, {Federici}, {Feng}, {Finley}, {Finnegan}, {Fortson}, {Furniss},
  {Galante}, {Gall}, {Gillanders}, {Godambe}, {Griffin}, {Grube}, {Gyuk},
  {Hanna}, {Holder}, {Huan}, {Hughes}, {Humensky}, {Kaaret}, {Karlsson},
  {Khassen}, {Kieda}, {Krawczynski}, {Krennrich}, {Lang}, {LeBohec}, {Lee},
  {Lyutikov}, {Madhavan}, {Maier}, {Majumdar}, {McArthur}, {McCann},
  {Moriarty}, {Mukherjee}, {Nelson}, {O'Faol{\'a}in de Bhr{\'o}ithe}, {Ong},
  {Orr}, {Otte}, {Park}, {Perkins}, {Pohl}, {Prokoph}, {Quinn}, {Ragan},
  {Reyes}, {Reynolds}, {Roache}, {Saxon}, {Schroedter}, {Sembroski}, {{\c
  S}ent{\"u}rk}, {Smith}, {Staszak}, {Telezhinsky}, {Te{\v s}i{\'c}},
  {Theiling}, {Thibadeau}, {Tsurusaki}, {Varlotta}, {Vincent}, {Vivier},
  {Wagner}, {Wakely}, {Weekes}, {Weinstein}, {Welsing}, {Williams}, {Zitzer},
  \& {Kondratiev}}]{2012ApJ...760..136A}
{Aliu}, E., {Archambault}, S., {Arlen}, T., {Aune}, T., {Beilicke}, M.,
  {Benbow}, W., {Bouvier}, A., {Buckley}, J.~H., {Bugaev}, V., {Byrum}, K.,
  {Cesarini}, A., {Ciupik}, L., {Collins-Hughes}, E., {Connolly}, M.~P., {Cui},
  W., {Dickherber}, R., {Duke}, C., {Dumm}, J., {Falcone}, A., {Federici}, S.,
  {Feng}, Q., {Finley}, J.~P., {Finnegan}, G., {Fortson}, L., {Furniss}, A.,
  {Galante}, N., {Gall}, D., {Gillanders}, G.~H., {Godambe}, S., {Griffin}, S.,
  {Grube}, J., {Gyuk}, G., {Hanna}, D., {Holder}, J., {Huan}, H., {Hughes}, G.,
  {Humensky}, T.~B., {Kaaret}, P., {Karlsson}, N., {Khassen}, Y., {Kieda}, D.,
  {Krawczynski}, H., {Krennrich}, F., {Lang}, M.~J., {LeBohec}, S., {Lee}, K.,
  {Lyutikov}, M., {Madhavan}, A.~S., {Maier}, G., {Majumdar}, P., {McArthur},
  S., {McCann}, A., {Moriarty}, P., {Mukherjee}, R., {Nelson}, T.,
  {O'Faol{\'a}in de Bhr{\'o}ithe}, A., {Ong}, R.~A., {Orr}, M., {Otte}, A.~N.,
  {Park}, N., {Perkins}, J.~S., {Pohl}, M., {Prokoph}, H., {Quinn}, J.,
  {Ragan}, K., {Reyes}, L.~C., {Reynolds}, P.~T., {Roache}, E., {Saxon}, D.~B.,
  {Schroedter}, M., {Sembroski}, G.~H., {{\c S}ent{\"u}rk}, G.~D., {Smith},
  A.~W., {Staszak}, D., {Telezhinsky}, I., {Te{\v s}i{\'c}}, G., {Theiling},
  M., {Thibadeau}, S., {Tsurusaki}, K., {Varlotta}, A., {Vincent}, S.,
  {Vivier}, M., {Wagner}, R.~G., {Wakely}, S.~P., {Weekes}, T.~C., {Weinstein},
  A., {Welsing}, R., {Williams}, D.~A., {Zitzer}, B., \& {Kondratiev}, V. 2012,
  \apj, 760, 136

\bibitem[{{Atoyan}(1999)}]{1999A&A...346L..49A}
{Atoyan}, A.~M. 1999, \aap, 346, L49

\bibitem[{{Bazin} {et~al.}(2009){Bazin}, {Palanque-Delabrouille}, {Rich},
  {Ruhlmann-Kleider}, {Aubourg}, {Le Guillou}, {Astier}, {Balland}, {Basa},
  {Carlberg}, {Conley}, {Fouchez}, {Guy}, {Hardin}, {Hook}, {Howell}, {Pain},
  {Perrett}, {Pritchet}, {Regnault}, {Sullivan}, {Antilogus}, {Arsenijevic},
  {Baumont}, {Fabbro}, {Le Du}, {Lidman}, {Mouchet}, {Mour{\~a}o}, \&
  {Walker}}]{2009A&A...499..653B}
{Bazin}, G., {Palanque-Delabrouille}, N., {Rich}, J., {Ruhlmann-Kleider}, V.,
  {Aubourg}, E., {Le Guillou}, L., {Astier}, P., {Balland}, C., {Basa}, S.,
  {Carlberg}, R.~G., {Conley}, A., {Fouchez}, D., {Guy}, J., {Hardin}, D.,
  {Hook}, I.~M., {Howell}, D.~A., {Pain}, R., {Perrett}, K., {Pritchet}, C.~J.,
  {Regnault}, N., {Sullivan}, M., {Antilogus}, P., {Arsenijevic}, V.,
  {Baumont}, S., {Fabbro}, S., {Le Du}, J., {Lidman}, C., {Mouchet}, M.,
  {Mour{\~a}o}, A., \& {Walker}, E.~S. 2009, \aap, 499, 653

\bibitem[{{Beskin} {et~al.}(2015){Beskin}, {Chernov}, {Gwinn}, \&
  {Tchekhovskoy}}]{2015SSRv..191..207B}
{Beskin}, V.~S., {Chernov}, S.~V., {Gwinn}, C.~R., \& {Tchekhovskoy}, A.~A.
  2015, \ssr, 191, 207

\bibitem[{{Bilous} {et~al.}(2012){Bilous}, {McLaughlin}, {Kondratiev}, \&
  {Ransom}}]{2012ApJ...749...24B}
{Bilous}, A.~V., {McLaughlin}, M.~A., {Kondratiev}, V.~I., \& {Ransom}, S.~M.
  2012, \apj, 749, 24

\bibitem[{{Burgay} {et~al.}(2003){Burgay}, {D'Amico}, {Possenti}, {Manchester},
  {Lyne}, {Joshi}, {McLaughlin}, {Kramer}, {Sarkissian}, {Camilo}, {Kalogera},
  {Kim}, \& {Lorimer}}]{2003Natur.426..531B}
{Burgay}, M., {D'Amico}, N., {Possenti}, A., {Manchester}, R.~N., {Lyne},
  A.~G., {Joshi}, B.~C., {McLaughlin}, M.~A., {Kramer}, M., {Sarkissian},
  J.~M., {Camilo}, F., {Kalogera}, V., {Kim}, C., \& {Lorimer}, D.~R. 2003,
  \nat, 426, 531

\bibitem[{{Caleb} {et~al.}(2016){Caleb}, {Flynn}, {Bailes}, {Barr}, {Bateman},
  {Bhandari}, {Campbell-Wilson}, {Green}, {Hunstead}, {Jameson}, {Jankowski},
  {Keane}, {Ravi}, {van Straten}, \& {Venkataraman
  Krishnan}}]{2016arXiv160102444C}
{Caleb}, M., {Flynn}, C., {Bailes}, M., {Barr}, E.~D., {Bateman}, T.,
  {Bhandari}, S., {Campbell-Wilson}, D., {Green}, A.~J., {Hunstead}, R.~W.,
  {Jameson}, A., {Jankowski}, F., {Keane}, E.~F., {Ravi}, V., {van Straten},
  W., \& {Venkataraman Krishnan}, V. 2016, ArXiv e-prints

\bibitem[{{Camelio} {et~al.}(2016){Camelio}, {Gualtieri}, {Pons}, \&
  {Ferrari}}]{2016arXiv160102945C}
{Camelio}, G., {Gualtieri}, L., {Pons}, J.~A., \& {Ferrari}, V. 2016, ArXiv
  e-prints

\bibitem[{{Camilo} {et~al.}(2006){Camilo}, {Ransom}, {Halpern}, {Reynolds},
  {Helfand}, {Zimmerman}, \& {Sarkissian}}]{2006Natur.442..892C}
{Camilo}, F., {Ransom}, S.~M., {Halpern}, J.~P., {Reynolds}, J., {Helfand},
  D.~J., {Zimmerman}, N., \& {Sarkissian}, J. 2006, \nat, 442, 892

\bibitem[{{Champion} {et~al.}(2015){Champion}, {Petroff}, {Kramer}, {Keith},
  {Bailes}, {Barr}, {Bates}, {Bhat}, {Burgay}, {Burke-Spolaor}, {Flynn},
  {Jameson}, {Johnston}, {Ng}, {Levin}, {Possenti}, {Stappers}, {van Straten},
  {Tiburzi}, \& {Lyne}}]{2015arXiv151107746C}
{Champion}, D.~J., {Petroff}, E., {Kramer}, M., {Keith}, M.~J., {Bailes}, M.,
  {Barr}, E.~D., {Bates}, S.~D., {Bhat}, N.~D.~R., {Burgay}, M.,
  {Burke-Spolaor}, S., {Flynn}, C.~M.~L., {Jameson}, A., {Johnston}, S., {Ng},
  C., {Levin}, L., {Possenti}, A., {Stappers}, B.~W., {van Straten}, W.,
  {Tiburzi}, C., \& {Lyne}, A.~G. 2015, ArXiv e-prints

\bibitem[{{Connor} {et~al.}(2016{\natexlab{a}}){Connor}, {Lin}, {Masui},
  {Oppermann}, {Pen}, {Peterson}, {Roman}, \& {Sievers}}]{2016arXiv160207292C}
{Connor}, L., {Lin}, H.-H., {Masui}, K., {Oppermann}, N., {Pen}, U.-L.,
  {Peterson}, J.~B., {Roman}, A., \& {Sievers}, J. 2016{\natexlab{a}}, ArXiv
  e-prints

\bibitem[{{Connor} {et~al.}(2016{\natexlab{b}}){Connor}, {Sievers}, \&
  {Pen}}]{2016MNRAS.458L..19C}
{Connor}, L., {Sievers}, J., \& {Pen}, U.-L. 2016{\natexlab{b}}, \mnras, 458,
  L19

\bibitem[{{Cordes} \& {Wasserman}(2016)}]{2016MNRAS.457..232C}
{Cordes}, J.~M. \& {Wasserman}, I. 2016, \mnras, 457, 232

\bibitem[{{Dahlen} {et~al.}(2012){Dahlen}, {Strolger}, {Riess}, {Mattila},
  {Kankare}, \& {Mobasher}}]{2012ApJ...757...70D}
{Dahlen}, T., {Strolger}, L.-G., {Riess}, A.~G., {Mattila}, S., {Kankare}, E.,
  \& {Mobasher}, B. 2012, \apj, 757, 70

\bibitem[{{Falcke} \& {Rezzolla}(2014)}]{2014A&A...562A.137F}
{Falcke}, H. \& {Rezzolla}, L. 2014, \aap, 562, A137

\bibitem[{{Gaensler} {et~al.}(2005){Gaensler}, {Kouveliotou}, {Gelfand},
  {Taylor}, {Eichler}, {Wijers}, {Granot}, {Ramirez-Ruiz}, {Lyubarsky},
  {Hunstead}, {Campbell-Wilson}, {van der Horst}, {McLaughlin}, {Fender},
  {Garrett}, {Newton-McGee}, {Palmer}, {Gehrels}, \& {Woods}}]{Gaensler1806}
{Gaensler}, B.~M., {Kouveliotou}, C., {Gelfand}, J.~D., {Taylor}, G.~B.,
  {Eichler}, D., {Wijers}, R.~A.~M.~J., {Granot}, J., {Ramirez-Ruiz}, E.,
  {Lyubarsky}, Y.~E., {Hunstead}, R.~W., {Campbell-Wilson}, D., {van der
  Horst}, A.~J., {McLaughlin}, M.~A., {Fender}, R.~P., {Garrett}, M.~A.,
  {Newton-McGee}, K.~J., {Palmer}, D.~M., {Gehrels}, N., \& {Woods}, P.~M.
  2005, \nat, 434, 1104

\bibitem[{{Hankins} \& {Eilek}(2007)}]{2007ApJ...670..693H}
{Hankins}, T.~H. \& {Eilek}, J.~A. 2007, \apj, 670, 693

\bibitem[{{Hankins} {et~al.}(2003){Hankins}, {Kern}, {Weatherall}, \&
  {Eilek}}]{2003Natur.422..141H}
{Hankins}, T.~H., {Kern}, J.~S., {Weatherall}, J.~C., \& {Eilek}, J.~A. 2003,
  \nat, 422, 141

\bibitem[{{Jauncey} {et~al.}(2001){Jauncey}, {Kedziora-Chudczer}, {Lovell},
  {Macquart}, {Nicolson}, {Perley}, {Reynolds}, {Tzioumis}, {Wieringa}, \&
  {Bignall}}]{2001Ap&SS.278...87J}
{Jauncey}, D.~L., {Kedziora-Chudczer}, L., {Lovell}, J.~E.~J., {Macquart},
  J.-P., {Nicolson}, G.~D., {Perley}, R.~A., {Reynolds}, J.~E., {Tzioumis},
  A.~K., {Wieringa}, M.~H., \& {Bignall}, H.~E. 2001, \apss, 278, 87

\bibitem[{{Karastergiou} {et~al.}(2015){Karastergiou}, {Chennamangalam},
  {Armour}, {Williams}, {Mort}, {Dulwich}, {Salvini}, {Magro}, {Roberts},
  {Serylak}, {Doo}, {Bilous}, {Breton}, {Falcke}, {Grie{\ss}meier}, {Hessels},
  {Keane}, {Kondratiev}, {Kramer}, {van Leeuwen}, {Noutsos}, {Os{\l}owski},
  {Sobey}, {Stappers}, \& {Weltevrede}}]{2015MNRAS.452.1254K}
{Karastergiou}, A., {Chennamangalam}, J., {Armour}, W., {Williams}, C., {Mort},
  B., {Dulwich}, F., {Salvini}, S., {Magro}, A., {Roberts}, S., {Serylak}, M.,
  {Doo}, A., {Bilous}, A.~V., {Breton}, R.~P., {Falcke}, H., {Grie{\ss}meier},
  J.-M., {Hessels}, J.~W.~T., {Keane}, E.~F., {Kondratiev}, V.~I., {Kramer},
  M., {van Leeuwen}, J., {Noutsos}, A., {Os{\l}owski}, S., {Sobey}, C.,
  {Stappers}, B.~W., \& {Weltevrede}, P. 2015, \mnras, 452, 1254

\bibitem[{{Keane} {et~al.}(2016){Keane}, {Johnston}, {Bhandari}, {Barr},
  {Bhat}, {Burgay}, {Caleb}, {Flynn}, {Jameson}, {Kramer}, {Petroff},
  {Possenti}, {van Straten}, {Bailes}, {Burke-Spolaor}, {Eatough}, {Stappers},
  {Totani}, {Honma}, {Furusawa}, {Hattori}, {Morokuma}, {Niino}, {Sugai},
  {Terai}, {Tominaga}, {Yamasaki}, {Yasuda}, {Allen}, {Cooke}, {Jencson},
  {Kasliwal}, {Kaplan}, {Tingay}, {Williams}, {Wayth}, {Chandra}, {Perrodin},
  {Berezina}, {Mickaliger}, \& {Bassa}}]{2016Natur.530..453K}
{Keane}, E.~F., {Johnston}, S., {Bhandari}, S., {Barr}, E., {Bhat}, N.~D.~R.,
  {Burgay}, M., {Caleb}, M., {Flynn}, C., {Jameson}, A., {Kramer}, M.,
  {Petroff}, E., {Possenti}, A., {van Straten}, W., {Bailes}, M.,
  {Burke-Spolaor}, S., {Eatough}, R.~P., {Stappers}, B.~W., {Totani}, T.,
  {Honma}, M., {Furusawa}, H., {Hattori}, T., {Morokuma}, T., {Niino}, Y.,
  {Sugai}, H., {Terai}, T., {Tominaga}, N., {Yamasaki}, S., {Yasuda}, N.,
  {Allen}, R., {Cooke}, J., {Jencson}, J., {Kasliwal}, M.~M., {Kaplan}, D.~L.,
  {Tingay}, S.~J., {Williams}, A., {Wayth}, R., {Chandra}, P., {Perrodin}, D.,
  {Berezina}, M., {Mickaliger}, M., \& {Bassa}, C. 2016, \nat, 530, 453

\bibitem[{{Keane} {et~al.}(2012){Keane}, {Stappers}, {Kramer}, \&
  {Lyne}}]{2012MNRAS.425L..71K}
{Keane}, E.~F., {Stappers}, B.~W., {Kramer}, M., \& {Lyne}, A.~G. 2012, \mnras,
  425, L71

\bibitem[{{Kulkarni} {et~al.}(2014){Kulkarni}, {Ofek}, {Neill}, {Zheng}, \&
  {Juric}}]{2014ApJ...797...70K}
{Kulkarni}, S.~R., {Ofek}, E.~O., {Neill}, J.~D., {Zheng}, Z., \& {Juric}, M.
  2014, \apj, 797, 70

\bibitem[{{Lai} {et~al.}(2001){Lai}, {Chernoff}, \&
  {Cordes}}]{2001ApJ...549.1111L}
{Lai}, D., {Chernoff}, D.~F., \& {Cordes}, J.~M. 2001, \apj, 549, 1111

\bibitem[{{Lang}(1999)}]{1999acfp.book.....L}
{Lang}, K.~R. 1999, {Astrophysical formulae}

\bibitem[{{Li} {et~al.}(2016){Li}, {Huang}, {Zhang}, {Li}, \&
  {Li}}]{2016arXiv160206099L}
{Li}, L., {Huang}, Y., {Zhang}, Z., {Li}, D., \& {Li}, B. 2016, ArXiv e-prints

\bibitem[{{Lorimer} {et~al.}(1993){Lorimer}, {Bailes}, {Dewey}, \&
  {Harrison}}]{1993MNRAS.263..403L}
{Lorimer}, D.~R., {Bailes}, M., {Dewey}, R.~J., \& {Harrison}, P.~A. 1993,
  \mnras, 263, 403

\bibitem[{{Lorimer} {et~al.}(2007){Lorimer}, {Bailes}, {McLaughlin},
  {Narkevic}, \& {Crawford}}]{2007Sci...318..777L}
{Lorimer}, D.~R., {Bailes}, M., {McLaughlin}, M.~A., {Narkevic}, D.~J., \&
  {Crawford}, F. 2007, Science, 318, 777

\bibitem[{{Luan} \& {Goldreich}(2014)}]{2014ApJ...785L..26L}
{Luan}, J. \& {Goldreich}, P. 2014, \apjl, 785, L26

\bibitem[{{Lundgren} {et~al.}(1995){Lundgren}, {Cordes}, {Ulmer}, {Matz},
  {Lomatch}, {Foster}, \& {Hankins}}]{1995ApJ...453..433L}
{Lundgren}, S.~C., {Cordes}, J.~M., {Ulmer}, M., {Matz}, S.~M., {Lomatch}, S.,
  {Foster}, R.~S., \& {Hankins}, T. 1995, \apj, 453, 433

\bibitem[{{Lyubarsky}(2014)}]{2014MNRAS.442L...9L}
{Lyubarsky}, Y. 2014, \mnras, 442, L9

\bibitem[{{Lyutikov}(2002)}]{lyutikovradiomagnetar}
{Lyutikov}, M. 2002, \apjl, 580, L65

\bibitem[{{Lyutikov}(2007)}]{2007MNRAS.381.1190L}
---. 2007, \mnras, 381, 1190

\bibitem[{{Lyutikov}(2015)}]{2015MNRAS.447.1407L}
---. 2015, \mnras, 447, 1407

\bibitem[{{Lyutikov} {et~al.}(1999){Lyutikov}, {Blandford}, \&
  {Machabeli}}]{1999MNRAS.305..338L}
{Lyutikov}, M., {Blandford}, R.~D., \& {Machabeli}, G. 1999, \mnras, 305, 338

\bibitem[{{Macquart} \& {Johnston}(2015)}]{2015MNRAS.451.3278M}
{Macquart}, J.-P. \& {Johnston}, S. 2015, \mnras, 451, 3278

\bibitem[{{Manchester} {et~al.}(2005){Manchester}, {Hobbs}, {Teoh}, \&
  {Hobbs}}]{ATNF}
{Manchester}, R.~N., {Hobbs}, G.~B., {Teoh}, A., \& {Hobbs}, M. 2005, \aj, 129,
  1993

\bibitem[{{Masui} {et~al.}(2015){Masui}, {Lin}, {Sievers}, {Anderson}, {Chang},
  {Chen}, {Ganguly}, {Jarvis}, {Kuo}, {Li}, {Liao}, {McLaughlin}, {Pen},
  {Peterson}, {Roman}, {Timbie}, {Voytek}, \& {Yadav}}]{2015Natur.528..523M}
{Masui}, K., {Lin}, H.-H., {Sievers}, J., {Anderson}, C.~J., {Chang}, T.-C.,
  {Chen}, X., {Ganguly}, A., {Jarvis}, M., {Kuo}, C.-Y., {Li}, Y.-C., {Liao},
  Y.-W., {McLaughlin}, M., {Pen}, U.-L., {Peterson}, J.~B., {Roman}, A.,
  {Timbie}, P.~T., {Voytek}, T., \& {Yadav}, J.~K. 2015, \nat, 528, 523

\bibitem[{{Melrose}(1992)}]{1992RSPTA.341..105M}
{Melrose}, D.~B. 1992, Philosophical Transactions of the Royal Society of
  London Series A, 341, 105

\bibitem[{{Melrose}(1995)}]{1995JApA...16..137M}
---. 1995, Journal of Astrophysics and Astronomy, 16, 137

\bibitem[{{Melrose} \& {Gedalin}(1999)}]{1999ApJ...521..351M}
{Melrose}, D.~B. \& {Gedalin}, M.~E. 1999, \apj, 521, 351

\bibitem[{{Mickaliger} {et~al.}(2012){Mickaliger}, {McLaughlin}, {Lorimer},
  {Langston}, {Bilous}, {Kondratiev}, {Lyutikov}, {Ransom}, \&
  {Palliyaguru}}]{2012ApJ...760...64M}
{Mickaliger}, M.~B., {McLaughlin}, M.~A., {Lorimer}, D.~R., {Langston}, G.~I.,
  {Bilous}, A.~V., {Kondratiev}, V.~I., {Lyutikov}, M., {Ransom}, S.~M., \&
  {Palliyaguru}, N. 2012, \apj, 760, 64

\bibitem[{{Moffett} \& {Hankins}(1996)}]{1996ApJ...468..779M}
{Moffett}, D.~A. \& {Hankins}, T.~H. 1996, \apj, 468, 779

\bibitem[{{Nomoto} \& {Kondo}(1991)}]{1991ApJ...367L..19N}
{Nomoto}, K. \& {Kondo}, Y. 1991, \apjl, 367, L19

\bibitem[{{Pen} \& {Connor}(2015)}]{2015ApJ...807..179P}
{Pen}, U.-L. \& {Connor}, L. 2015, \apj, 807, 179

\bibitem[{{Petroff} {et~al.}(2015{\natexlab{a}}){Petroff}, {Bailes}, {Barr},
  {Barsdell}, {Bhat}, {Bian}, {Burke-Spolaor}, {Caleb}, {Champion}, {Chandra},
  {Da Costa}, {Delvaux}, {Flynn}, {Gehrels}, {Greiner}, {Jameson}, {Johnston},
  {Kasliwal}, {Keane}, {Keller}, {Kocz}, {Kramer}, {Leloudas}, {Malesani},
  {Mulchaey}, {Ng}, {Ofek}, {Perley}, {Possenti}, {Schmidt}, {Shen},
  {Stappers}, {Tisserand}, {van Straten}, \& {Wolf}}]{2015MNRAS.447..246P}
{Petroff}, E., {Bailes}, M., {Barr}, E.~D., {Barsdell}, B.~R., {Bhat},
  N.~D.~R., {Bian}, F., {Burke-Spolaor}, S., {Caleb}, M., {Champion}, D.,
  {Chandra}, P., {Da Costa}, G., {Delvaux}, C., {Flynn}, C., {Gehrels}, N.,
  {Greiner}, J., {Jameson}, A., {Johnston}, S., {Kasliwal}, M.~M., {Keane},
  E.~F., {Keller}, S., {Kocz}, J., {Kramer}, M., {Leloudas}, G., {Malesani},
  D., {Mulchaey}, J.~S., {Ng}, C., {Ofek}, E.~O., {Perley}, D.~A., {Possenti},
  A., {Schmidt}, B.~P., {Shen}, Y., {Stappers}, B., {Tisserand}, P., {van
  Straten}, W., \& {Wolf}, C. 2015{\natexlab{a}}, \mnras, 447, 246

\bibitem[{{Petroff} {et~al.}(2016){Petroff}, {Barr}, {Jameson}, {Keane},
  {Bailes}, {Kramer}, {Morello}, {Tabbara}, \& {van
  Straten}}]{2016arXiv160103547P}
{Petroff}, E., {Barr}, E.~D., {Jameson}, A., {Keane}, E.~F., {Bailes}, M.,
  {Kramer}, M., {Morello}, V., {Tabbara}, D., \& {van Straten}, W. 2016, ArXiv
  e-prints

\bibitem[{{Petroff} {et~al.}(2015{\natexlab{b}}){Petroff}, {Johnston}, {Keane},
  {van Straten}, {Bailes}, {Barr}, {Barsdell}, {Burke-Spolaor}, {Caleb},
  {Champion}, {Flynn}, {Jameson}, {Kramer}, {Ng}, {Possenti}, \&
  {Stappers}}]{2015MNRAS.454..457P}
{Petroff}, E., {Johnston}, S., {Keane}, E.~F., {van Straten}, W., {Bailes}, M.,
  {Barr}, E.~D., {Barsdell}, B.~R., {Burke-Spolaor}, S., {Caleb}, M.,
  {Champion}, D.~J., {Flynn}, C., {Jameson}, A., {Kramer}, M., {Ng}, C.,
  {Possenti}, A., \& {Stappers}, B.~W. 2015{\natexlab{b}}, \mnras, 454, 457

\bibitem[{{Petroff} {et~al.}(2014){Petroff}, {van Straten}, {Johnston},
  {Bailes}, {Barr}, {Bates}, {Bhat}, {Burgay}, {Burke-Spolaor}, {Champion},
  {Coster}, {Flynn}, {Keane}, {Keith}, {Kramer}, {Levin}, {Ng}, {Possenti},
  {Stappers}, {Tiburzi}, \& {Thornton}}]{2014ApJ...789L..26P}
{Petroff}, E., {van Straten}, W., {Johnston}, S., {Bailes}, M., {Barr}, E.~D.,
  {Bates}, S.~D., {Bhat}, N.~D.~R., {Burgay}, M., {Burke-Spolaor}, S.,
  {Champion}, D., {Coster}, P., {Flynn}, C., {Keane}, E.~F., {Keith}, M.~J.,
  {Kramer}, M., {Levin}, L., {Ng}, C., {Possenti}, A., {Stappers}, B.~W.,
  {Tiburzi}, C., \& {Thornton}, D. 2014, \apjl, 789, L26

\bibitem[{{Phinney}(1991)}]{1991ApJ...380L..17P}
{Phinney}, E.~S. 1991, \apjl, 380, L17

\bibitem[{{Popov} \& {Stappers}(2007)}]{2007A&A...470.1003P}
{Popov}, M.~V. \& {Stappers}, B. 2007, \aap, 470, 1003

\bibitem[{{Popov} \& {Postnov}(2010)}]{2010vaoa.conf..129P}
{Popov}, S.~B. \& {Postnov}, K.~A. 2010, in Evolution of Cosmic Objects through
  their Physical Activity, ed. H.~A. {Harutyunian}, A.~M. {Mickaelian}, \&
  Y.~{Terzian}, 129--132

\bibitem[{{Popov} \& {Turolla}(2012)}]{2012Ap&SS.341..457P}
{Popov}, S.~B. \& {Turolla}, R. 2012, \apss, 341, 457

\bibitem[{{Radhakrishnan} \& {Cooke}(1969)}]{1969ApL.....3..225R}
{Radhakrishnan}, V. \& {Cooke}, D.~J. 1969, \aplett, 3, 225

\bibitem[{{Rane} {et~al.}(2016){Rane}, {Lorimer}, {Bates}, {McMann},
  {McLaughlin}, \& {Rajwade}}]{2016MNRAS.455.2207R}
{Rane}, A., {Lorimer}, D.~R., {Bates}, S.~D., {McMann}, N., {McLaughlin},
  M.~A., \& {Rajwade}, K. 2016, \mnras, 455, 2207

\bibitem[{{Rowlinson} {et~al.}(2016){Rowlinson}, {Bell}, {Murphy}, {Trott},
  {Hurley-Walker}, {Johnston}, {Tingay}, {Kaplan}, {Carbone}, {Hancock},
  {Feng}, {Offringa}, {Bernardi}, {Bowman}, {Briggs}, {Cappallo}, {Deshpande},
  {Gaensler}, {Greenhill}, {Hazelton}, {Johnston-Hollitt}, {Lonsdale},
  {McWhirter}, {Mitchell}, {Morales}, {Morgan}, {Oberoi}, {Ord}, {Prabu},
  {Udaya Shankar}, {Srivani}, {Subrahmanyan}, {Wayth}, {Webster}, {Williams},
  \& {Williams}}]{2016arXiv160207544R}
{Rowlinson}, A., {Bell}, M.~E., {Murphy}, T., {Trott}, C.~M., {Hurley-Walker},
  N., {Johnston}, S., {Tingay}, S.~J., {Kaplan}, D.~L., {Carbone}, D.,
  {Hancock}, P.~J., {Feng}, L., {Offringa}, A.~R., {Bernardi}, G., {Bowman},
  J.~D., {Briggs}, F., {Cappallo}, R.~J., {Deshpande}, A.~A., {Gaensler},
  B.~M., {Greenhill}, L.~J., {Hazelton}, B.~J., {Johnston-Hollitt}, M.,
  {Lonsdale}, C.~J., {McWhirter}, S.~R., {Mitchell}, D.~A., {Morales}, M.~F.,
  {Morgan}, E., {Oberoi}, D., {Ord}, S.~M., {Prabu}, T., {Udaya Shankar}, N.,
  {Srivani}, K.~S., {Subrahmanyan}, R., {Wayth}, R.~B., {Webster}, R.~L.,
  {Williams}, A., \& {Williams}, C.~L. 2016, ArXiv e-prints

\bibitem[{{Sincell} \& {Krolik}(1992)}]{1992ApJ...395..553S}
{Sincell}, M.~W. \& {Krolik}, J.~H. 1992, \apj, 395, 553

\bibitem[{{Soglasnov}(2007)}]{2007whsn.conf...68S}
{Soglasnov}, V. 2007, in WE-Heraeus Seminar on Neutron Stars and Pulsars 40
  years after the Discovery, ed. W.~{Becker} \& H.~H. {Huang}, 68

\bibitem[{{Soglasnov} {et~al.}(2004){Soglasnov}, {Popov}, {Bartel}, {Cannon},
  {Novikov}, {Kondratiev}, \& {Altunin}}]{2004ApJ...616..439S}
{Soglasnov}, V.~A., {Popov}, M.~V., {Bartel}, N., {Cannon}, W., {Novikov},
  A.~Y., {Kondratiev}, V.~I., \& {Altunin}, V.~I. 2004, \apj, 616, 439

\bibitem[{{Spitler} {et~al.}(2014){Spitler}, {Cordes}, {Hessels}, {Lorimer},
  {McLaughlin}, {Chatterjee}, {Crawford}, {Deneva}, {Kaspi}, {Wharton},
  {Allen}, {Bogdanov}, {Brazier}, {Camilo}, {Freire}, {Jenet},
  {Karako-Argaman}, {Knispel}, {Lazarus}, {Lee}, {van Leeuwen}, {Lynch},
  {Ransom}, {Scholz}, {Siemens}, {Stairs}, {Stovall}, {Swiggum},
  {Venkataraman}, {Zhu}, {Aulbert}, \& {Fehrmann}}]{2014ApJ...790..101S}
{Spitler}, L.~G., {Cordes}, J.~M., {Hessels}, J.~W.~T., {Lorimer}, D.~R.,
  {McLaughlin}, M.~A., {Chatterjee}, S., {Crawford}, F., {Deneva}, J.~S.,
  {Kaspi}, V.~M., {Wharton}, R.~S., {Allen}, B., {Bogdanov}, S., {Brazier}, A.,
  {Camilo}, F., {Freire}, P.~C.~C., {Jenet}, F.~A., {Karako-Argaman}, C.,
  {Knispel}, B., {Lazarus}, P., {Lee}, K.~J., {van Leeuwen}, J., {Lynch}, R.,
  {Ransom}, S.~M., {Scholz}, P., {Siemens}, X., {Stairs}, I.~H., {Stovall}, K.,
  {Swiggum}, J.~K., {Venkataraman}, A., {Zhu}, W.~W., {Aulbert}, C., \&
  {Fehrmann}, H. 2014, \apj, 790, 101

\bibitem[{{Spitler} {et~al.}(2016){Spitler}, {Scholz}, {Hessels}, {Bogdanov},
  {Brazier}, {Camilo}, {Chatterjee}, {Cordes}, {Crawford}, {Deneva}, {Ferdman},
  {Freire}, {Kaspi}, {Lazarus}, {Lynch}, {Madsen}, {McLaughlin}, {Patel},
  {Ransom}, {Seymour}, {Stairs}, {Stappers}, {van Leeuwen}, \&
  {Zhu}}]{2016arXiv160300581S}
{Spitler}, L.~G., {Scholz}, P., {Hessels}, J.~W.~T., {Bogdanov}, S., {Brazier},
  A., {Camilo}, F., {Chatterjee}, S., {Cordes}, J.~M., {Crawford}, F.,
  {Deneva}, J., {Ferdman}, R.~D., {Freire}, P.~C.~C., {Kaspi}, V.~M.,
  {Lazarus}, P., {Lynch}, R., {Madsen}, E.~C., {McLaughlin}, M.~A., {Patel},
  C., {Ransom}, S.~M., {Seymour}, A., {Stairs}, I.~H., {Stappers}, B.~W., {van
  Leeuwen}, J., \& {Zhu}, W.~W. 2016, ArXiv e-prints

\bibitem[{{Tendulkar} {et~al.}(2016){Tendulkar}, {Kaspi}, \&
  {Patel}}]{2016arXiv160202188T}
{Tendulkar}, S.~P., {Kaspi}, V.~M., \& {Patel}, C. 2016, ArXiv e-prints

\bibitem[{{Thornton} {et~al.}(2013){Thornton}, {Stappers}, {Bailes},
  {Barsdell}, {Bates}, {Bhat}, {Burgay}, {Burke-Spolaor}, {Champion}, {Coster},
  {D'Amico}, {Jameson}, {Johnston}, {Keith}, {Kramer}, {Levin}, {Milia}, {Ng},
  {Possenti}, \& {van Straten}}]{2013Sci...341...53T}
{Thornton}, D., {Stappers}, B., {Bailes}, M., {Barsdell}, B., {Bates}, S.,
  {Bhat}, N.~D.~R., {Burgay}, M., {Burke-Spolaor}, S., {Champion}, D.~J.,
  {Coster}, P., {D'Amico}, N., {Jameson}, A., {Johnston}, S., {Keith}, M.,
  {Kramer}, M., {Levin}, L., {Milia}, S., {Ng}, C., {Possenti}, A., \& {van
  Straten}, W. 2013, Science, 341, 53

\bibitem[{{Vink}(2008)}]{2008AdSpR..41..503V}
{Vink}, J. 2008, Advances in Space Research, 41, 503

\bibitem[{{Wang} \& {Gotthelf}(1998)}]{1998ApJ...509L.109W}
{Wang}, Q.~D. \& {Gotthelf}, E.~V. 1998, \apjl, 509, L109

\bibitem[{{Williams} \& {Berger}(2016)}]{2016arXiv160208434W}
{Williams}, P.~K.~G. \& {Berger}, E. 2016, ArXiv e-prints

\bibitem[{{Wilson} \& {Rees}(1978)}]{1978MNRAS.185..297W}
{Wilson}, D.~B. \& {Rees}, M.~J. 1978, \mnras, 185, 297

\end{thebibliography}

    \appendix

   \section{$f(\dot{E})$ distribution from  $f(B,P)$}
    \label{fBP} 
 Let's assume that at birth the distribution of \Bfs\ and periods is $f(B,P)$. 
We can parametrize the  spin-down power as 
\be
\dot{E} =  \dot{E}_0  \left( {B\over B_0} \right)^2 \left( {P\over P_0} \right)^{-4}
\ee
where $B_0$, $P_0$   and $\dot{E}_0$ are some fiducial values of the \Bf,  period and spin-down power.
For a given  $\dot{E}$ we have 
\ba &&
 B/B_0= (\Phi/2 )  (P/P_0) ^{2}
\nn &&
\Phi = 2  \sqrt{\dot{E}/ \dot{E}_0}
\ea
(function $\Phi $ is proportional to the total  electric potential).

\begin{figure}[htbp]
\begin{center}
\includegraphics[width=0.99\linewidth]{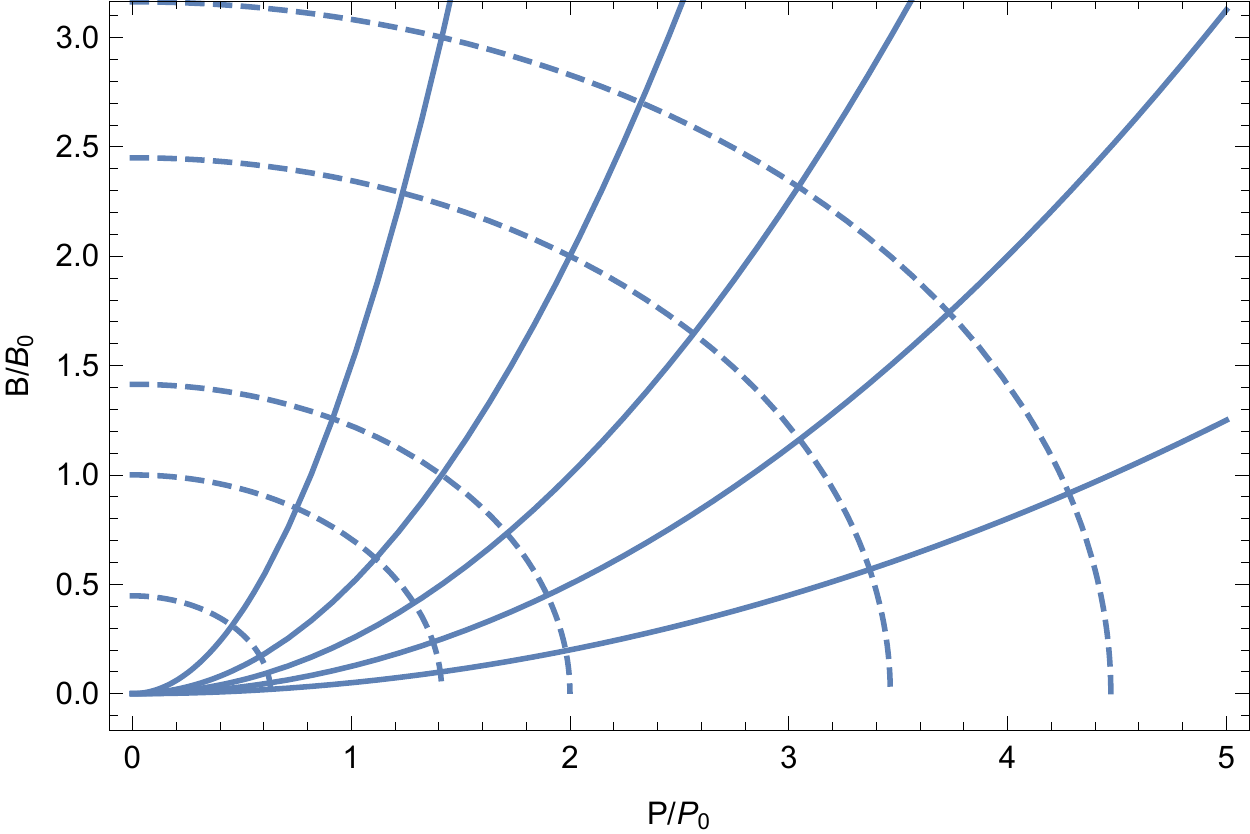}
\caption{Lines of constant $\sqrt{\dot{E}}\equiv  \Phi$ (solid) and orthogonal curves (lines of constant $\Psi$).}
\label{BPEdot}
\end{center}
\end{figure}

We can introduce a coordinate  $\Psi $ orthogonal to the  electric potential
\ba &&
\Psi={1\over 4} \left( 2 \left( {B\over B_0} \right)^2 +  \left( {P\over P_0} \right)^2\right)
\nn &&
P/P_0 = \sqrt{\frac{\sqrt{8 \Phi^2 \Psi+1}-1}{\Phi^2}}
\nn &&
B/B_0=\frac{\sqrt{8 \Phi^2 \Psi+1}-1}{2 \Phi},
\label{BP}
  \ea
  see Fig. \ref{BPEdot}.
(If \Bf\ remains constant and $P_0$ and $B_0$ are the initial values, a given pulsar follows a line $\Psi = (1+ \Phi)/(2 \Phi)$ starting from a point $\Phi_0 =2$ and $\Psi_0 =3/4$.)

The Jacobian of the transformation $\{B,P\} \rightarrow  \{\Phi,\Psi\}$ is 
  \be
  J = - \sqrt{ \sqrt{ 1+ 8 \Phi^2 \Psi} - 1 \over \Phi^2 ( 1+ 8 \Phi^2 \Psi) } 
  \ee
  To find distribution in potential (and spin-down power)  $f(\Phi) d \Phi   = f(\sqrt{\dot{E}}) \dot{E} ^{-1/2}  d   \dot{E} /2$ we need to integrate
  \be
  f(\Phi,\Psi)= f(B,T) J
  \label{fab}
  \ee
  (where \Bf\ and period are expressed  by Eq. (\ref{BP}) )  over $\Psi$ (from zero to infinity). 

For example, for a log-normal injection  distribution in both \Bf\ and period, with mean $P_0$ and $B_0$ and dispersions $\sigma_{P,B}$,
the resulting $\dot{E}$ distribution is also log-normal. 
 \ba &&
f(\dot{E}) ={  e^{- \ln^2 (\dot{E}/\dot{E}_0) /2 \sigma^2}\over 2 \sqrt{2 \pi} \sigma} \, {1\over  \dot{E} } 
\nn &&
\sigma = \sqrt{ \sigma_B^2 + 4 \sigma_P^2}
\nn &&
\bar{\dot{E}} = e^{ 2 \sigma^2} {\dot{E}}_0
\ea
Log-normal distribution closely resembles the  $ 1/\dot{E}$ power-law over a broad range of parameters.

\end{document}